\documentclass[a4paper,12pt]{article}
\pdfoutput=1
\usepackage{hyperref}
\usepackage{pstricks}
\usepackage{color}
\usepackage{amssymb,amsmath,bm,bbm}
\usepackage{epsf,epsfig}
\usepackage{afterpage}
\usepackage{longtable}
\usepackage{cite}
\usepackage{latexsym, mathrsfs}
\usepackage{graphicx,graphics,epsfig,epstopdf,subfigure,psfrag}% Include figure files
\usepackage{url}
\usepackage{paralist}
\usepackage{bbold}
\newcommand{\be}{\begin{equation}}
\newcommand{\ee}{\end{equation}}
\newcommand{\bea}{\begin{eqnarray}}
\newcommand{\eea}{\end{eqnarray}}
\newcommand{\bec}{\begin{center}}
\newcommand{\eec}{\end{center}}
\newcommand{\nn}{\nonumber}
\newcommand{\dd}{\displaystyle}

\begin{document}

\begin{flushright}{BARI-TH/15-695} \end{flushright}
\medskip

\begin{center}
{\LARGE\bf
\boldmath{On thermalization   of a boost-invariant \\ \vspace{0.5cm}   non-Abelian plasma
}}\\[0.8 cm]
{\bf L.~Bellantuono$^{a,b}$, P.~Colangelo$^{b}$, F.~De~Fazio$^{b}$,
F.~Giannuzzi$^{a}$
 \\[0.5 cm]}
{\small
$^a$Dipartimento di Fisica, Universit\`a  di Bari,via Orabona 4, I-70126 Bari, Italy\\
$^b$INFN, Sezione di Bari, via Orabona 4, I-70126 Bari, Italy\\}
\end{center}

\vskip0.61cm

\abstract{Using a holographic method, we further investigate the relaxation towards the hydrodynamic regime of a boost-invariant non-Abelian plasma taken out-of-equilibrium.  In the dual description, the system is driven out-of-equilibrium by   boundary sourcing,
a deformation of the boundary metric, 
as proposed by Chesler and Yaffe. The effects of  several deformation profiles on  the bulk geometry are investigated by the analysis of the corresponding solutions of the Einstein equations.  The time of restoration of the hydrodynamic regime is investigated:
 setting the effective temperature of the system at the  end of the boundary quenching to  $T_{eff}(\tau^*)=500$ MeV, the hydrodynamic regime  is reached after a lapse of time of ${\cal O}$(1 fm/c). }

\vspace*{2cm}

\thispagestyle{empty}
\newpage

\section{Introduction}
The possibility of employing the gauge/gravity duality correspondence \cite{Maldacena:1997re,Witten:1998qj, Gubser:1998bc}  to the analysis of far-from-equilibrium processes in strongly coupled systems 
\footnote{For a review see \cite{Hubeny:2010ry}.} has enlarged the realm of nonperturbative phenomena to which this  theoretical method can be applied in a quite controllable way, while traditional  approaches are less effective.  Among the different systems, of particular interest is the one produced in ultrarelativistic heavy ion collisions (HI), as those taking place at RHIC and at LHC. The features of this system, for time scales larger than about 1 fm/c, seem to be well reproduced in a  hydrodynamic setup involving a strongly coupled/low viscosity  fluid \cite{Heinz:2004pj,Shuryak:2014zxa}, a   framework allowing to correctly describe experimental observables  such as the  light hadron   transverse momentum spectra.
In the hydrodynamic scheme 
of crucial importance are, after  the pre-equilibrium regime  following the collisions,
the  conditions at the time when the hydrodynamic behavior sets up. In particular,  at this time the system stress-energy tensor $ T^\mu_\nu$ is important. We write it as
\be 
T^\mu_\nu=\frac{N_c^2}{2 \pi^2}\, diag(-\epsilon, p_\perp,p_\perp, p_\parallel) \,\,\, , \label{stress-en-tens}
 \ee 
in terms of     the system energy density $\epsilon$ and of  $p_\perp,  p_\parallel$ the  pressures  along one of the two transverse directions (with respect to the collision axis)   and in the longitudinal direction, respectively \footnote{ Along the paper, we  refer to energy density and pressures without considering  the  factor $\frac{N_c^2}{2 \pi^2}$.}. For example,  in approaches based on the idea of an initial state described by  color glass  condensate with a saturation scale $Q_s$ of the order of a few GeV, the 
initial stress-energy  tensor
 comes from  classical chromo-electric and chromo-magnetic fields and has  the form  
$ T^\mu_\nu \propto diag(-\epsilon, \epsilon, \epsilon,-\epsilon)$.  However, such an initial  condition has been shown to be unstable, as soon as one-loop corrections in the strong coupling constant 
%$g_s$ 
are switched on 
%and   $p_\perp\simeq p_\parallel$ is obtained already for quite small values  $g_s\simeq 0.5$,   sizably smaller than the value relevant in  HI collisions 
\cite{Gelis:2013rba}.

 In the analysis  presented  in our study,  a prominent role is  therefore played by the stress-energy tensor, which is needed to understand the features of the equilibrium regime,  the  time needed to reach equilibrium,  and the properties of the dynamics driving the system towards  equilibrium. 
Questions  to be answered are, for a system driven out-of-the-equilibrium, whether the late-time  state is  described by hydrodynamics, and whether the elapsed time needed to
 reach  equilibrium is related to  the dynamics during the out-of-equilibrium state. These issues concerning the transient process from a far-from-equilibrium to a  hydrodynamic regime  can be faced in the holographic approach  \cite{CasalderreySolana:2011us}, using as a guideline the conjectured
correspondence   between  a supersymmetric, conformal field theory defined in a four dimensional (4D) space-time and a  gravity theory in a  higher dimensional space-time,  a five dimensional  Anti de-Sitter space (AdS$_5$) times a  five dimensional sphere  S$^5$ \cite{Maldacena:1997re}.  The  gauge theory is defined on the 4D  boundary of AdS$_5$, and the connection has been extended to the  case of  nonlinear fluid dynamics \cite{Bhattacharyya:2008jc}.
 Non-equilibrium  can be studied by solving in the bulk  the 5D Einstein equations for the metric subject to suitable time-dependent boundary conditions. 
Information about the  late-time   regime can be obtained computing various invariants. For example, in Ref.\cite{Janik:2005zt}   the square of the Riemann tensor $\Re^2$ (the  Kretschmann scalar)  has been studied  as an expansion in inverse powers of the proper time up to second order, finding that, in the asymptotic $\tau \to \infty$ regime, the scalar $\Re^2$  is free of singularities in  the perfect fluid case.    The boundary theory stress-energy tensor results  from the  solutions of the Einstein equations in the bulk.

Analyses in the case of boost-invariant fluid \cite{Chesler:2008hg,Chesler:2009cy},  for some choices of the profile distortion, have shown the formation of  a horizon  in the bulk,  and have given access to   the components of the boundary theory stress-energy tensor. 
The investigations have been extended to  samples of  initial states  characterized by different values of the components of the stress-energy tensor, without  reference to the mechanism producing each state
\cite{Heller:2011ju,Heller:2012je,Heller:2012km,Wu:2011yd,Heller:2013oxa,Jankowski:2014lna}.
 
As formulated in   \cite{Chesler:2008hg,Chesler:2009cy}, the initial off-equilibrium states can be thought as being produced by external  sources distorting the boundary metric for short time intervals. We  focus on this issue, with a study of external distortion profiles,   step-shaped or   short-duration sequences   repeating themselves with different intensities, with the aim of describing  phenomena where a small number of collisions takes place before the system starts evolving to  thermal equilibrium.  This study in the boost-invariant framework is useful for more involved cases where  a lower symmetry is assumed for the system, for example in shock-wave collisions \cite{Grumiller:2008va,Chesler:2010bi,vanderSchee:2013pia,Chesler:2015wra}.

There are several reasons to study  complex distortion shapes in the boundary metric.  The bulk Einstein equations are nonlinear, and  different distorsions produce unrelated responses. The resulting effective temperature and  entropy density follow different profiles as a consequence of the boundary sourcing.  Moreover, it is interesting to investigate whether the onset of the hydrodynamic regime is related to the distortion curve. Finally, these analyses have to   face problems concerning the stability of the solution of GR equations with different boundary conditions.

% This paper deals with  such issues. 
 The plan of the present paper, in which a few of the above issues are examined,   is the following.  In Section \ref{sec2} we describe how a system is taken out-of-equilibrium through the distortion of the boundary metric, as done in \cite{Chesler:2008hg,Chesler:2009cy}. We discuss a suitable procedure for
solving the resulting 5D Einstein equations, providing  a few details of the algorithm and the issue of the stability of the solutions. In Section \ref{sec3} we describe the profiles of the boundary geometry distortion we have chosen to study,  with  the motivations. In Section \ref{sec4} we discuss our findings, in particular concerning  the equilibration time.   The conclusions are presented in the last Section. Details dealing with  the calculation of the boundary stress-energy tensor $T^\mu_\nu$ are collected in  appendix \ref{appA},  while in appendix \ref{numerics}  we discuss some aspects of the numerical algorithm.

\section{Distorting the boundary}\label{sec2}
The main idea, as  in \cite{Chesler:2008hg,Chesler:2009cy}, is to study the effects of a time-dependent deformation of the metric of the 4D boundary. As a consequence of the deformation, gravitational radiation is produced and propagates in the 5D bulk,   and a  black hole  is formed together with its horizon. In this way,  one can  trade  the study of the reach of  equilibrium in the 4D boundary  theory for the analysis  of the time evolution of the geometry in the 5D bulk.

We set the 4D coordinates $x^\mu=(x^0,x^1,x^2,x^3)$,   identifying  $x^3=x_\parallel$ with the axis in the direction of collisions and along which  the plasma expands. Boost-invariance along that axis is imposed, together with translation invariance and $O(2)$ rotation invariance  in the orthogonal plane    $x_\perp=\{x^1,\,x^2 \}$. In terms of  the proper time $\tau$ and  rapidity $y$, given by  $x^0=\tau \cosh y$, $x_\parallel=\tau \sinh y$,  the 4D Minkowski line element  is  given by
$ds^2=-d\tau^2+dx_\perp^2+\tau^2 dy^2$. 

A distortion of the boundary metric, leaving the spatial three-volume unchanged and respecting the  imposed symmetries, can be obtained through a function   $\gamma(\tau)$  which encodes information about a deformation:
\be
ds^2=-d\tau^2+e^{\gamma(\tau)} dx_\perp^2+\tau^2e^{-2\gamma(\tau)} dy^2 \,\,. \label{metric4D}
\ee
The space-time with  the metric (\ref{metric4D}) is considered as the boundary  of the 5D space-time in which the gravity dual  is defined. We adopt 5D Eddington-Finkelstein  coordinates,
with $r$ the extra-dimension coordinate and  the boundary  reached  for $r \to \infty$. In general, the 5D bulk metric can be chosen as
\be
ds^2=2 dr d\tau-A d\tau^2+ \Sigma^2 e^B dx_\perp^2+ \Sigma^2 e^{-2B}dy^2  \,\,\, , \label{metric5D}
\ee
with the  functions $A$, $\Sigma$ and $B$  only depending on $r$ and $\tau$ to respect  the adopted symmetries. The behavior of these functions against  the deformation allows us to describe  how the bulk metric changes as a consequence of the distortion of the boundary.  With the metric (\ref{metric5D}), at fixed $x_\perp$ and rapidity $y$, infalling radial null geodesics correspond to  constant $\tau$, while outgoing radial null geodesics are obtained from $\dd \frac{d r}{d \tau}=\frac{A(r,\tau)}{2}$. Therefore, for a generic function $\xi(r,\tau)$ the derivatives 
\bea
\xi^\prime&=&\partial_r \xi
\label{der-r} \\
{\dot \xi}&=& \partial_\tau \xi+\frac{1}{2} A \partial_r \xi
\label{der-dot}
\eea
represent  directional derivatives along the infalling radial null geodesics and the outgoing radial null geodesics,   respectively. 
 
The 5D metric (\ref{metric5D}) is the solution  of   Einstein's equations with negative cosmological constant. In terms of $A(r,\tau)$, $\Sigma(r, \tau)$ and  $B(r,\tau)$   the equations can be rephrased as  \cite{Chesler:2009cy}:
\bea
&&\Sigma ({\dot \Sigma})^\prime +2 \Sigma^\prime {\dot \Sigma}-2 \Sigma^2=0  \label{ein1} \\
&& \Sigma ({\dot B})^\prime+\frac{3}{2} \left(\Sigma^\prime {\dot B}+B^\prime {\dot \Sigma}\right)=0  \label{ein2}\\
&& A^{\prime \prime} +3 B^\prime {\dot B} -12 \frac{\Sigma^\prime {\dot \Sigma} }{\Sigma^2}+4=0\label{ein3} \\
&& {\ddot \Sigma}+\frac{1}{2} \left( {\dot B} ^2 \Sigma -A^\prime  {\dot \Sigma} \right) =0\label{ein4} \\
&& \Sigma^{\prime \prime}+\frac{1}{2} B^{\prime 2} \Sigma =0 \label{ein5}\,\,. \eea
The  five equations (\ref{ein1})-(\ref{ein5})  can be considered  as three dynamical and two constraint equations. 
The condition that the metric (\ref{metric5D}) produces the 4D metric  (\ref{metric4D}) at the boundary, for $r \to \infty$,  constrains  the large $r$ behavior of  $A(r,\tau)$, $\Sigma(r, \tau)$ and  $B(r,\tau)$:
\bea
&&\frac{A(r,\tau)-r^2 }{r}\xrightarrow[r \to \infty]{ } 0 \,\,, \label{large-r2} \\
&& \frac{\Sigma(r,\tau)}{r} \xrightarrow[r \to \infty]{ }  \tau^{\frac{1}{3}} \,\, , \label{large-r0} \\
&&B(r,\tau)\xrightarrow[r \to \infty]{ } -\frac{2}{3} \log \tau +\gamma(\tau)  \,\, .\label{large-r1} 
\eea
The invariance of (\ref{metric5D}) under the  diffeomorphism $r \to r + \lambda(\tau)$  has  been exploited to impose the large $r$ condition for $A$.

We  switch on  the distortion of the boundary metric at the initial time $\tau=\tau_i$,  starting from the  AdS$_5$ bulk metric:
\be
ds^2= 2 dr d\tau + r^2 \left[ -d \tau^2+ d x_\perp^2 + \left( \tau+\frac{1}{r} \right)^2 dy^2 \right]   \,\,\, . \label{AdS5}
\ee
As discussed in \cite{Heller:2012je}, taking the limit  $r \to \infty$ and $\tau \to 0$ with the Eddington-Finkelstein coordinates, Eq.~(\ref{AdS5}),  is  ambiguous. For this reason, we set $\tau_i >0$.  The  initial conditions for $A$, $\Sigma$ and $B$ are therefore:
\bea
A(r,\tau_i) &=& A_{ini}(r)=r^2  \,\,\, , \label{ini2} \\
\Sigma (r,\tau_i)&=&\Sigma_{ini}(r)=r \left(\tau_i+\frac{1}{r} \right)^{1/3}  \,\,\, , \label{ini0}  \\ 
B(r,\tau_i)&=& B_{ini}(r)=-\frac{2}{3} \log \left( \tau_i+\frac{1}{r} \right)  \,\,\, . \label{ini1} 
\eea

For $r \to \infty$,  the Einstein's equations can be solved starting from  the relations (\ref{large-r2})-(\ref{large-r1}).
As suggested in \cite{Chesler:2008hg,Chesler:2009cy},   for    $A(r,\tau)$, $\Sigma(r, \tau)$ and   $B(r,\tau)$  the  large-$r$ expansions can be written as
\bea
A_{asy}(r,\tau) &=& \sum_{n=0} \left[a_n(\tau)+\alpha_n(\tau) \log r \right] r^{2-n} \,\,  ,  \label{asy2} \\
\Sigma_{asy}(r,\tau)&=& \sum_{n=0} \left[s_n(\tau)+\sigma_n(\tau) \log r \right] r^{1-n}  \,\, , \label{asy0}\\
B_{asy}(r,\tau) &=& \sum_{n=0} \left[b_n(\tau)+\beta_n(\tau) \log r \right] r^{-n}  \,\,\, . \label{asy1} 
\eea
The conditions  (\ref{large-r2})-(\ref{large-r1})  fix the  $n=0$ coefficients in  (\ref{asy2})-(\ref{asy1}),  as well as those for $n=1$ in  $A_{asy}$. The other coefficients can be determined imposing that Eqs.~(\ref{ein1})-(\ref{ein3})  are satisfied by the functions (\ref{asy2})-(\ref{asy1}), but this   leaves two  coefficients undetermined,  $a_4(\tau)$ and $b_4(\tau)$. 
From  Eq.~(\ref{ein5}) a relation follows between  $a_4$ and $b_4$:
\be
b_4(\tau) =  \frac{ 1296 \tau^5  a_4^\prime (\tau) +1728 \tau^4 a_4(\tau)+G(\tau) }{1728 \tau^4 \left(-2+3\tau \gamma^\prime(\tau)\right)} \,\,, \label{eq:a4b4}
\ee
with  the primes denoting   derivatives with respect to $\tau$, and the function $G(\tau)$ expressed in terms of   $\gamma^{(n)}(\tau)=\frac{d^n \gamma(\tau)}{d \tau^n}$:
\bea
G(\tau) &=&-576 +3618 \tau \gamma^\prime(\tau) -\tau^2\left(6903 \gamma^\prime(\tau)^2+1474 \gamma^{\prime \prime}(\tau)\right) \nn \\
&+&\tau^3 \left(4608 \gamma^\prime(\tau)^3+5199 \gamma^\prime(\tau) \gamma^{\prime \prime}(\tau)+468 \gamma^{(3)} (\tau) \right) \nn \\
&+&\tau^4 \left( -918  \gamma^\prime(\tau)^4-6480  \gamma^\prime(\tau)^2 \gamma^{\prime \prime}(\tau) +216 \gamma^{\prime \prime}(\tau)^2-540  \gamma^\prime(\tau)  \gamma^{(3)} (\tau)+306  \gamma^{(4)} (\tau) \right) \nn \\
&+&\tau^5 \left( 3240\gamma^\prime(\tau)^3  \gamma^{\prime \prime}(\tau)+162  \gamma^{\prime \prime}(\tau) \gamma^{(3)} (\tau)-459 \gamma^\prime(\tau)  \gamma^{(4)}(\tau) \right) \label{G} \,\,.
\eea

In our procedure, $a_4(\tau)$ and $b_4(\tau)$ are determined by Eq.~\eqref{eq:a4b4} and matching the computed functions $\Sigma$ and $B$ with their asymptotic expressions \eqref{asy0} and \eqref{asy1} for large  $r$. Using $a_4(\tau)$ and $b_4(\tau)$, important observables can be obtained, such as the components of the boundary stress-energy tensor $T^\mu_\nu$, as discussed in next sections.

The details of our method for solving Eqs.~\eqref{ein1}-\eqref{ein5}  are as follows.
In the set of functions $\Sigma$, $\dot\Sigma$, $B$, $\dot B$ and $A$, we treat a given function and its derivative (\ref{der-dot}) as independent.
At $\tau=\tau_i$, $\Sigma(r,\tau_i)$, $B(r,\tau_i)$ and $A(r,\tau_i)$ are known, Eqs.~(\ref{ini2})-(\ref{ini1}). To complete the  set of initial conditions, we compute $\dot\Sigma(r,\tau_i)$ and $\dot B(r,\tau_i)$  solving Eqs.~(\ref{ein1}) and (\ref{ein2}), respectively.

At $\tau>\tau_i$ the algorithm is organized in  steps.
\begin{enumerate}
 \item Using  the definition (\ref{der-dot}), we determine $\partial_\tau \Sigma(r,\tau_i)$ and $\partial_\tau B(r,\tau_i)$, which allow us to obtain $\Sigma$ and $B$ at the subsequent time $\tau_i+d\tau$, if $d\tau$ is small enough for the equations
$$\Sigma(r,\tau_i+d\tau)=\Sigma(r,\tau_i)+d\tau \partial_\tau \Sigma(r,\tau_i)$$
$$B(r,\tau_i+d\tau)=B(r,\tau_i)+d\tau \partial_\tau B(r,\tau_i)$$
 to be valid. 
 \item Once $\Sigma(r,\tau_i+d\tau)$ and $B(r,\tau_i+d\tau)$ are determined, we  compute $a_4(\tau_i+d\tau)$ and $b_4(\tau_i+d\tau)$. Indeed,  $a_4(\tau_i+d\tau)$  can be obtained  fitting   the computed functions $\Sigma(r,\tau_i+d\tau)$ and $B(r,\tau_i+d\tau)$ with their asymptotic expressions \eqref{asy0} and \eqref{asy1} in the range $r_{as}\leqslant r \leqslant r_{max}$  (later on in this section, it will be shown how to set  $r_{as})$. Then, we get $b_4(\tau_i+d\tau)$ from Eq.~\eqref{eq:a4b4}.
 \item $\dot \Sigma(r,\tau_i+d\tau)$, $\dot B(r,\tau_i+d\tau)$ and finally $A(r,\tau_i+d\tau)$  are computed solving  Eqs.~(\ref{ein1})-(\ref{ein3}); the boundary conditions are fixed using \eqref{asy0}, \eqref{asy1}, \eqref{asy2}, and    $a_4(\tau_i+d\tau)$ and $b_4(\tau_i+d\tau)$.
\end{enumerate}
With the functions known at $\tau_i+d\tau$, the cycle is repeated until a chosen final value of $\tau$ is reached. Typically, with the $\gamma$ functions considered in our analysis we use $d\tau=10^{-2}$.

At each $\tau$ step, our solution algorithm  requires three values of the bulk coordinate $r$: $r_{max}$ and $r_{min}$, which are the maximum and minimum value of $r$ used in the numerical calculation, and $r_{as}$, a value employed to determine $a_4(\tau)$ from the asymptotic expressions (\ref{asy2})-(\ref{asy1})  used in the range $[r_{as},r_{max}]$. Stability against variation of such parameters is used as a criterium to check the numerical results. Particularly relevant is the minimum value $r_{min}$, required to perform an excision in the bulk coordinates. This value is set in the following way. At fixed $\tau$,  for each one of  Eqs.~(\ref{ein1})-(\ref{ein5}) we construct the function ${\cal R}_i(r,\tau)$  defined as the absolute value of the l.h.s. of the  equation divided by the  sum of the absolute value of its addendi (see appendix \ref{numerics} for definitions).
At $\tau=\tau_i$, we choose a small initial value of $r_{min}$; then, at each $\tau$, $r_{min}$ is increased by steps of size typically $0.005$, until the differential equations (\ref{ein1})-(\ref{ein3}) and one constraint equation (\ref{ein5}) are   fulfilled, namely by requiring the condition ${\cal R}_i(r_{min},\tau)<0.03$ for $i=6,7,8,10$.
As a result, the bulk excision  is always beyond the apparent horizon. In  $r_{min}$,  the function $A(r_{min},\tau)$ becomes negative but not large, hence possible singularities beyond the horizon are avoided. 
A similar procedure is carried out for $r_{as}$. We initially set $r_{as}$, and then we gradually increase it until the condition ${\cal R}_i(r_{as},\tau)<0.01$ for $i=6,7,8,10$ is satisfied.  
Therefore, at this stage, we monitor the accuracy of the three differential equations  and one constraint equation in $r_{min}$ and $r_{as}$. At the end of the procedure,  we check that the condition ${\cal R}_i\le 0.01$ is fulfilled for all  Eqs.~(\ref{ein1})-(\ref{ein5}), i.e. $i=6-10$, and in the whole $(r,\tau)$ range. We find that this is indeed the case except for   few tiny regions. In the largest part of the $(r,\,\tau)$ plane deviations from zero are smaller than ${\cal O}(10^{-4})$, as discussed  in appendix  \ref{numerics}. 
During  the numerical evaluation,   no corrections are required to account for the finite value of the time step.  In spite of the intricacy of the boundary conditions, stable solutions are  found.\footnote{Other methods for the numerical solution of analogous General Relativity  equations in the case of initial value problems are described in  \cite{Chesler:2013lia} and in references therein.} 

\section{Models for the distortion of the boundary geometry}\label{sec3}
We are interested in investigating deformations of the boundary metric characterized by different duration, intensity and structure.  We set the generic  form for $\gamma(\tau)$:
\be
\gamma(\tau) =w \left[ Tanh \left(\frac{\tau -\tau_0}{\eta} \right) \right]^7 \,+\sum_{j=1}^{N}\gamma_j(\tau,\tau_{0,j})  \label{profile}
\ee
with
\be
\gamma_j(\tau,\tau_{0,j}) = c_j f_j(\tau,\tau_{0,j})^6 e^{-1/f_j(\tau,\tau_{0,j})} \Theta\left(1-\frac{(\tau-\tau_{0,j})^2}{\Delta_j^2}\right)\,\, \label{profile1} 
\ee
and 
\be
f_j(\tau,\tau_{0,j})= 1- \frac{(\tau-\tau_{0,j})^2}{\Delta_j^2} \,\,\, .\label{profile2}
\ee
This expression generalizes the one used in \cite{Chesler:2009cy}, and represents sequences of $N$ "pulses",  each one of intensity $c_j$ and duration proportional to $\Delta_j$, with the possibility of superimposing a smooth deformation, of intensity $w$,  which asymptotically  changes the scales of the coordinates. After their generation, pulses travel in the longitudinal direction $x_\parallel$, driving the boundary  state out-of-equilibrium. The response of the $5D$ bulk geometry to this deformation describes how equilibrium is reached at late times.
We choose three different models, $\cal A$,   $\cal B$ and $\cal C$,   setting the various parameters in Eqs.~(\ref{profile})-(\ref{profile2}).

Model $\cal A$  represents two pulses in the boundary metric. The parameters are set to
$w=0$ and $N=2$, 
$c_1 = 1$, $\Delta_{1,2}=1$, $\tau_{0,1}=\frac{5 }{4}\Delta_1$,
$c_2 = 2$. Moreover,   two different values of $\tau_{0,2}$:   $\tau_{0,2}=\frac{17}{4} \Delta_2$ and $\tau_{0,2}=\frac{9}{4} \Delta_2$ are considered, changing the time interval  between the pulses.  The final time of the distortion is  $\tau_f^{\cal A}=5.25$ and  $\tau_f^{\cal A}=3.25$ for the two cases, respectively, two quantities important for our discussion.

In model $\cal B$ we set
$w=\frac{2}{5}$, $\eta=1.2$, $\tau_0=0.25$, $N=1$, 
$c_1 = 1$, $\Delta_1=1$, $\tau_{0,1}=4 \Delta_1$, and the short pulse ends at  $\tau_f^{\cal B}=5$ while a tale of the distortion continues with $\tau$ and approaches a constant value.

Model $\cal C$ combines both the previous ones, and is obtained using
\bea
w&=& \frac{2}{5}, \,\,\,\,\, \eta=1.2 , \,\,\,\,\, \tau_0=0.25 \,\,, \nn \\
c_{0,1}&=& 0.5, \,\,\,\,\,
c_{0,2}= 1, \,\,\,\,\,
c_{0,3}= 0.35, \,\,\,\,\,
c_{0,4}= 0.3, \nn \\
\tau_{0,1}&=&  \tau_0+1.5, \,\,\,\,\,
\tau_{0,2}=  \tau_0+3.8, \,\,\,\,\,
\tau_{0,3}=  \tau_0+6.1, \,\,\,\,\,
\tau_{0,4}=  \tau_0+8.4, \nn \\
\Delta_j&=&0.8 \,\,\,\,\, ( j=1,2,3,4)\,\,. \nn 
\eea
For this model the last short pulse ends at $\tau_f^{\cal C}=9.45$.
As mentioned, to avoid in  the Eddington-Finkelstein  coordinates the ambiguity in the limits $r \to \infty$ and $\tau \to 0$,  in all cases the initial  time  is $\tau_i>0$.
The three profiles $\gamma(\tau)$ are depicted in  Figs.~\ref{Column2p}, \ref{ColumnGp} and \ref{Column4p}. 

The distortion profiles are chosen with the purpose of studying  different situations.
Model $\cal A$ is the  repetition,  in  the boundary metric, of two short pulses of different intensity,  Fig.~\ref{Column2p}.  
In particular, we consider two cases with different interval of time between the pulses, ranging from the condition of distant perturbations to the case of overlapping pulses.
This model can provide us with information on the horizon formation, 
in comparison with the case of  isolated pulse studied in  \cite{Chesler:2009cy}, and on the effects related to the time elapsed between the distortions. 
Model $\cal B$ has the purpose of studying a combination of two effects  with very different time scales, a short pulse and a slow continuum  asymptotically producing a rescaling  of the boundary coordinates.  Model $\cal C$, a combination of the previous cases,  accounts for the effects of several pulses with various intensities,  and is closer to  physical processes  driving systems out-of equilibrium  in relativistic heavy ion collisions. In all cases, the horizon formation and behavior  are investigated, together with  the various components the boundary stress-energy tensor, 
 looking at  the time when the hydrodynamic regime sets in.

 \section{Results and discussions}\label{sec4}
In the numerical calculation for the models previously described, the solutions of the bulk Einstein equations allow to obtain the apparent horizon and the event horizon, which define the system temperature  $T_{eff}$. The apparent horizon is  determined as the locus of points where the condition $\dot \Sigma(r_h, \tau)=0$ is fulfilled. 
\footnote{The apparent horizon is the outermost trapped null surface. 
To compute it, a foliation of spacetime at fixed times is considered. In each spacelike hypersurface, the apparent horizon is a closed surface such that \cite{york}:
\begin{equation}\label{eq:AppHor}
H=\gamma^{ab} \nabla_a s_b  -   K + s^a s^b K_{ab} =0 \,\,\, , \nn
\end{equation}
with  $\gamma_{ab}$ the induced metric and $K_{ab}$  the extrinsic curvature of the hypersurface, $K = g^{ab} K_{ab}$, and $s^a$ the outward-pointing spacelike unit vector, normal to the apparent horizon and tangent to the hypersurface. For the spacetime described by  Eq.~\eqref{metric5D} this corresponds to the condition $\dot\Sigma=0$.}
The event horizon, which separates causally disconnected regions,  is found drawing outgoing radial null geodesic curves   and looking for the critical curve which separates the  geodesics escaping towards the boundary from the ones plunging back in the bulk. 
From these curves, the function $r_h(\tau)$  ($r_h(\tau)$ being the position of the horizon at the proper time $\tau$) is determined, which defines  the temperature $T_{eff}(\tau)$. Asymptotically in proper time, the apparent horizon and the event horizon coincide. 
\footnote{Studies of the behavior of the horizons in the gravity dual of a boost-invariant flow can be found in \cite{Bhattacharyya:2008xc,Booth:2009ct}. }
From the solution, $\Sigma(r_h,\tau)^3$,  the area of each horizon per unit of rapidity, can be computed.

The three models share common features, which we discuss together with the   differences.  Before presenting this analysis, let us  recall a few results concerning the boundary stress-energy tensor $T^\mu_\nu$. 

As discussed in the Bjorken's  seminal paper  \cite{Bjorken:1982qr}, under the assumptions of homogeneity,  boost-invariance and invariance under rotations  in the transverse plane with respect to the fluid velocity, the fluid stress-energy tensor $T_{\mu}^{\nu}$ has well-defined properties, and its components  only depend on the proper time $\tau$, a manifestly boost-invariant condition. Imposing  $T_{\mu}^{\nu}$ conservation and  traceless conditions,  the tensor components can  be expressed in terms of a single function $f(\tau)$, that can be chosen to be  $-T_0^0$:  
\be
T_{\mu}^{\nu}=diag \left(-f(\tau),\,f(\tau)+\displaystyle{\frac{1}{2}}\tau f^\prime(\tau),\,f(\tau)+\displaystyle{\frac{1}{2}}\tau f^\prime(\tau),\, -f(\tau)-\tau f^\prime(\tau)\right) \,\,\,\ .
\ee
For a  perfect fluid,  the equation of state $\epsilon=3 p$,  with $p=p_\parallel=p_\perp$,   fixes the $\tau$ dependence: $\epsilon(\tau)=\displaystyle{\frac{const}{\tau^{4/3}}}$, stemming  from the relations
\bea
p_\parallel (\tau)&=&-\epsilon(\tau) -\tau \epsilon^\prime(\tau) \label{eq-ppar}\\
p_\perp (\tau) &=& \epsilon(\tau) +\frac{\tau}{2} \epsilon^\prime(\tau)  \label{eq-pperp}\,\,\, . 
\eea 
Deviations from the ideal behavior can be included as corrections taking  viscous effects into account,  in a  gradient expansion. On the gauge theory side, the gradient expansion corresponds to a late time expansion, with subleading terms identified with the dissipative ones in the hydrodynamical theory. 
 As shown in \cite{Janik:2005zt}, the function $f(\tau)$ deviates from $f(\tau)=const/\tau^{4/3}$  if the assumption of perfect fluid behavior is removed. For example, the case
 $\epsilon (\tau)\sim \displaystyle{\frac{1}{\tau^{s}}}$ corresponds to $p_\parallel \sim  \displaystyle{\frac{(s-1)}{\tau^s}}$ and $p_\perp \sim \displaystyle{\frac{(2-s)}{2\tau^s}}$.~\footnote{In \cite{Heller:2012je}  the case  $0 \le s < 4$ is studied.
}
Subleading corrections to the perfect fluid behavior  relate the energy density
\be
\epsilon(\tau)= \frac{3}{4} \pi^4 T_{eff}(\tau)^4  \label{Teff}
\ee
to the definition of an effective temperature $T_{eff}(\tau)$. The calculation in ${\cal N}=4$  SYM gives \cite{Heller:2007qt,Baier:2007ix,Heller:2012je}:
\bea
T_{eff}(\tau)&=&\frac{\Lambda}{(\Lambda \tau)^{1/3}} \Bigg[ 1-\frac{1}{6 \pi (\Lambda \tau)^{2/3}}+\frac{-1+\log 2}{36 \pi^2 (\Lambda \tau)^{4/3} } \nn \\
&+&\frac{-21+2\pi^2+51 \log 2 -24 (\log 2)^2}{1944 \pi^3 (\Lambda \tau)^2} + {\cal O}\left( \frac{1}{(\Lambda \tau)^{8/3}} \right )\Bigg] \,\,\ . \label{Teff1}
\eea
This leads to 
\bea
\epsilon(\tau)&=& \frac{3 \pi^4 \Lambda^4}{4 (\Lambda \tau)^{4/3} }\left[ 1-\frac{2c_1}{ (\Lambda \tau)^{2/3}}+\frac{c_2}{ (\Lambda \tau)^{4/3}} + {\cal O}\left( \frac{1}{(\Lambda \tau)^2} \right )\right] \label{hydroeps} \\
p_\parallel (\tau)&=&\frac{ \pi^4 \Lambda^4}{ 4(\Lambda \tau)^{4/3} } \left[ 1-\frac{6c_1}{ (\Lambda \tau)^{2/3}}+\frac{5c_2}{ (\Lambda \tau)^{4/3}} + {\cal O}\left( \frac{1}{(\Lambda \tau)^2} \right )\right] \label{hydroppar} \\
p_\perp (\tau) &=& \frac{ \pi^4 \Lambda^4}{ 4(\Lambda \tau)^{4/3} } \left[ 1-\frac{c_2}{ (\Lambda \tau)^{4/3}} + {\cal O}\left( \frac{1}{(\Lambda \tau)^2} \right ) \right] \,\,\, ,\label{hydropperp}
\eea
 and  to the pressure ratio and  anisotropy:
\bea
\frac{p_\parallel}{p_\perp}&=& 1-\frac{6c_1}{ (\Lambda \tau)^{2/3}}+\frac{6c_2}{ (\Lambda \tau)^{4/3}} + {\cal O}\left( \frac{1}{(\Lambda \tau)^2} \right) ,\label{pressure-ratio} \\
\frac{\Delta p}{\epsilon} &=& \frac{p_\perp-p_\parallel}{\epsilon} =2\left[ \frac{c_1}{ (\Lambda \tau)^{2/3}}+\frac{2c_1^2-c_2}{ (\Lambda \tau)^{4/3}} +{\cal O}\left( \frac{1}{(\Lambda \tau)^2} \right) \right] \,\,\, ,\label{pressure-anis}
\eea
with  $c_1=\displaystyle{\frac{1}{3 \pi}}$ and $c_2=\displaystyle{\frac{1+2 \log{2}}{18 \pi^2}}$. $\Lambda$ is a parameter to be determined for each one of the considered models.

Understanding the results of the calculation of the solutions of the 5D Einstein equations, hence, requires the computation of the boundary stress-energy tensor $T^\mu_\nu$ in Eq.~(\ref{stress-en-tens}) and the comparison with the proper time dependence following the previous equations.
 The various components of the  stress-energy tensor 
of the boundary theory  can be determined using the holographic renormalization procedure developed in \cite{deHaro:2000xn,Kinoshita:2008dq} and shortly described in appendix \ref{appA}. The  functions $a_4(\tau)$ and $b_4(\tau)$ in the expansions   (\ref{asy2})  and   (\ref{asy1}), computed through the Einstein equations, are required, and  the energy density and the longitudinal and parallel pressures are obtained in terms of such functions:
\bea
\epsilon&=& -\frac{3}{4} a_4 + \tilde \epsilon_\gamma \,\,\, , \label{eps-a4} \\
p_\perp&=& -\frac{1}{4} a_4 +b_4 +  \tilde  p_{\perp, \gamma} \,\,\, ,\label{pperp-a4}  \\
p_\parallel&=& -\frac{1}{4} a_4 -2 b_4 + \tilde  p_{\parallel, \gamma} \,\,\, . \label{ppar-a4} 
\eea
The functions  $\tilde \epsilon_\gamma(\tau)$,  $\tilde  p_{\perp,\gamma}(\tau)$ and $\tilde  p_{\parallel,\gamma}(\tau)$  are related to the distortion profile  $\gamma(\tau)$ in the boundary metric,  and  are reported in appendix~\ref{appA}.

\begin{figure}[t!]
\bec
\begin{tabular}{ll}
\includegraphics[width = 0.45\textwidth]{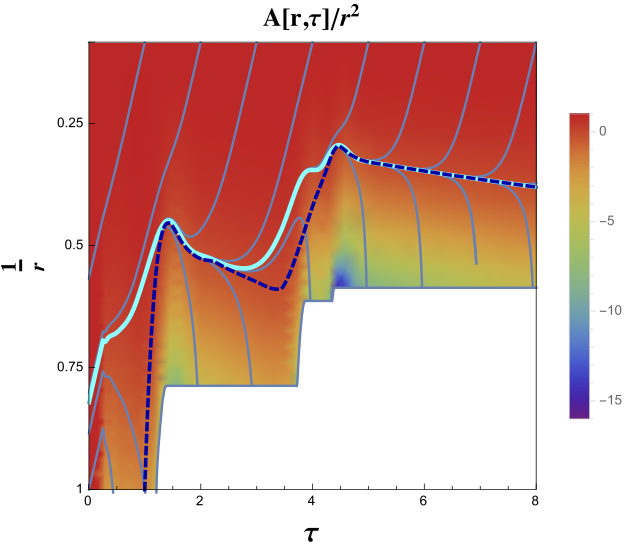}&\hspace*{-0.5cm}
\includegraphics[width = 0.5\textwidth]{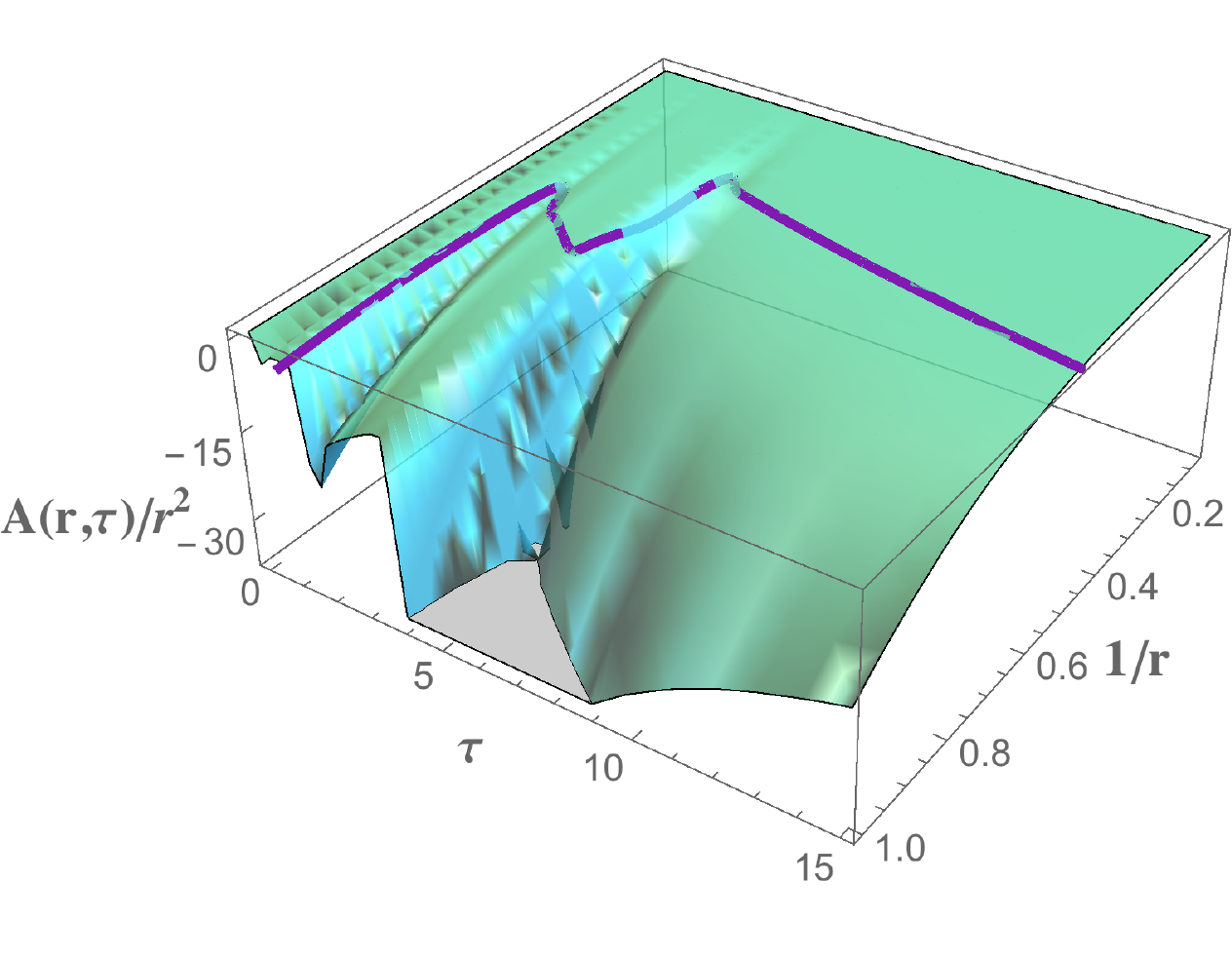}\\
\includegraphics[width = 0.45\textwidth]{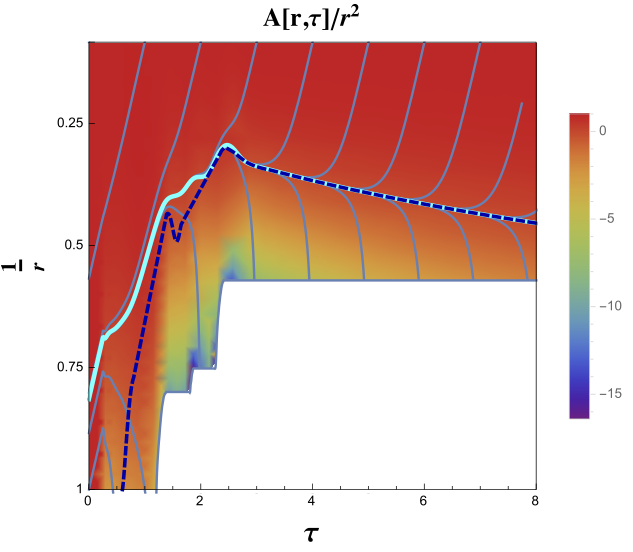}&\hspace*{-0.5cm}
\includegraphics[width = 0.5\textwidth]{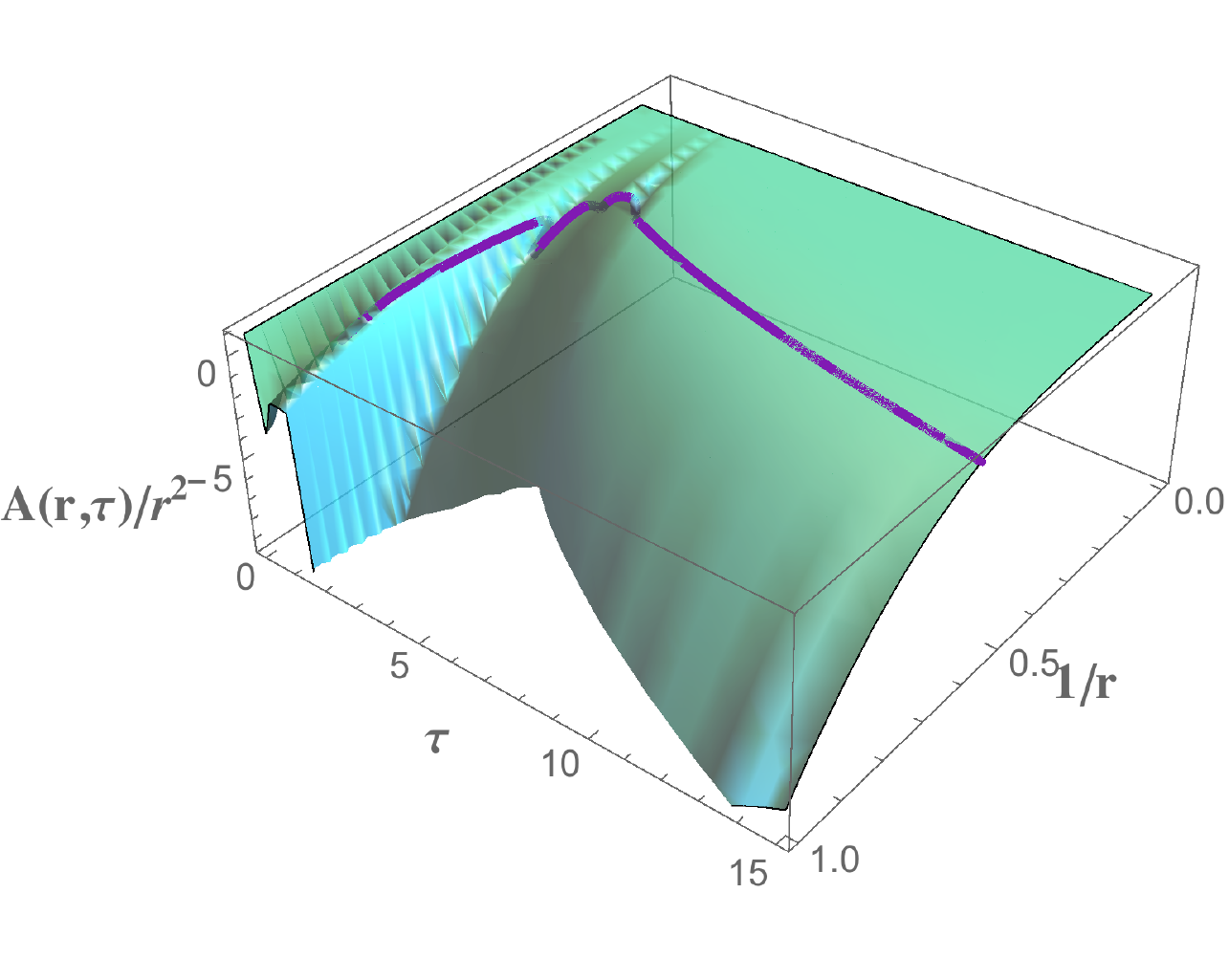}

\end{tabular}
\caption{\small {\bf Model $\cal A$.}  The left panel shows the function $A(r,\tau)/r^2$, vs $\tau$ and $1/r$,  obtained solving  the Einstein equations (\ref{ein1})-(\ref{ein5}) in the case of two distant (top panel) or close pulses (bottom panel)  in the boundary metric. The color bar indicates the values of the function. The gray lines are radial null outgoing geodesics,  the dashed  dark blue line is the apparent  horizon,  and the continuous cyan line is  the event horizon obtained as the critical geodesic. The  excision in the low-$r$ region, used in the calculation,  is  shown.  In the  right panels, the  3d representation of $A(r,\tau)/r^2$ is  displayed  together with the apparent horizon (continuous line).}\label{density-due-picchi}
\eec
\end{figure}

We can now discuss the three considered models, focusing on the effects of the boundary distortion and on the transient  from the far-from-equilibrium state to the hydrodynamic regime. \\

\noindent{\bf Model $\cal A$.}

\begin{figure}[t!]
\bec
\begin{tabular}{ll}
\includegraphics[width = 0.4\textwidth]{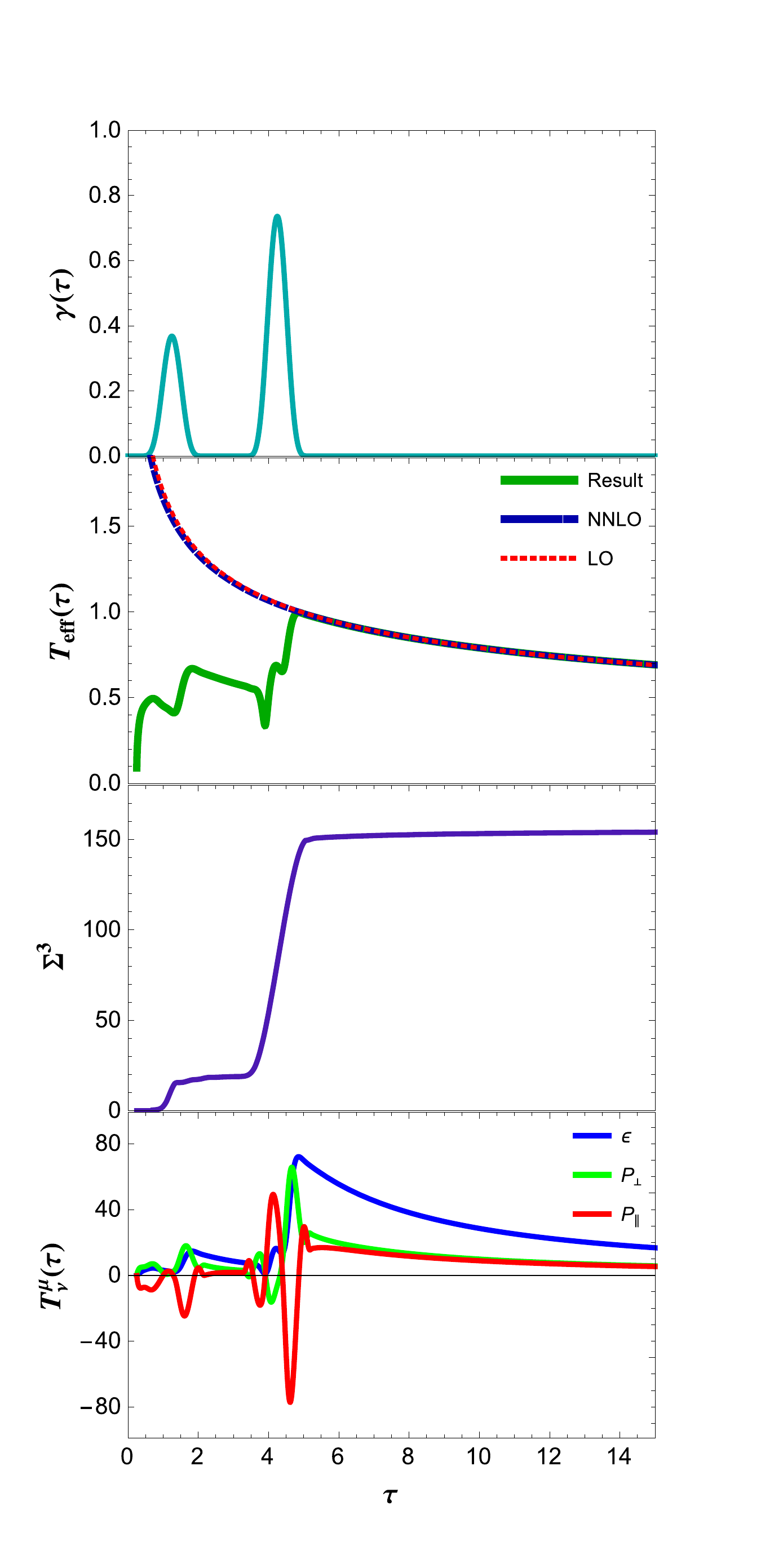}&\hspace*{-1.cm}
\includegraphics[width = 0.4\textwidth]{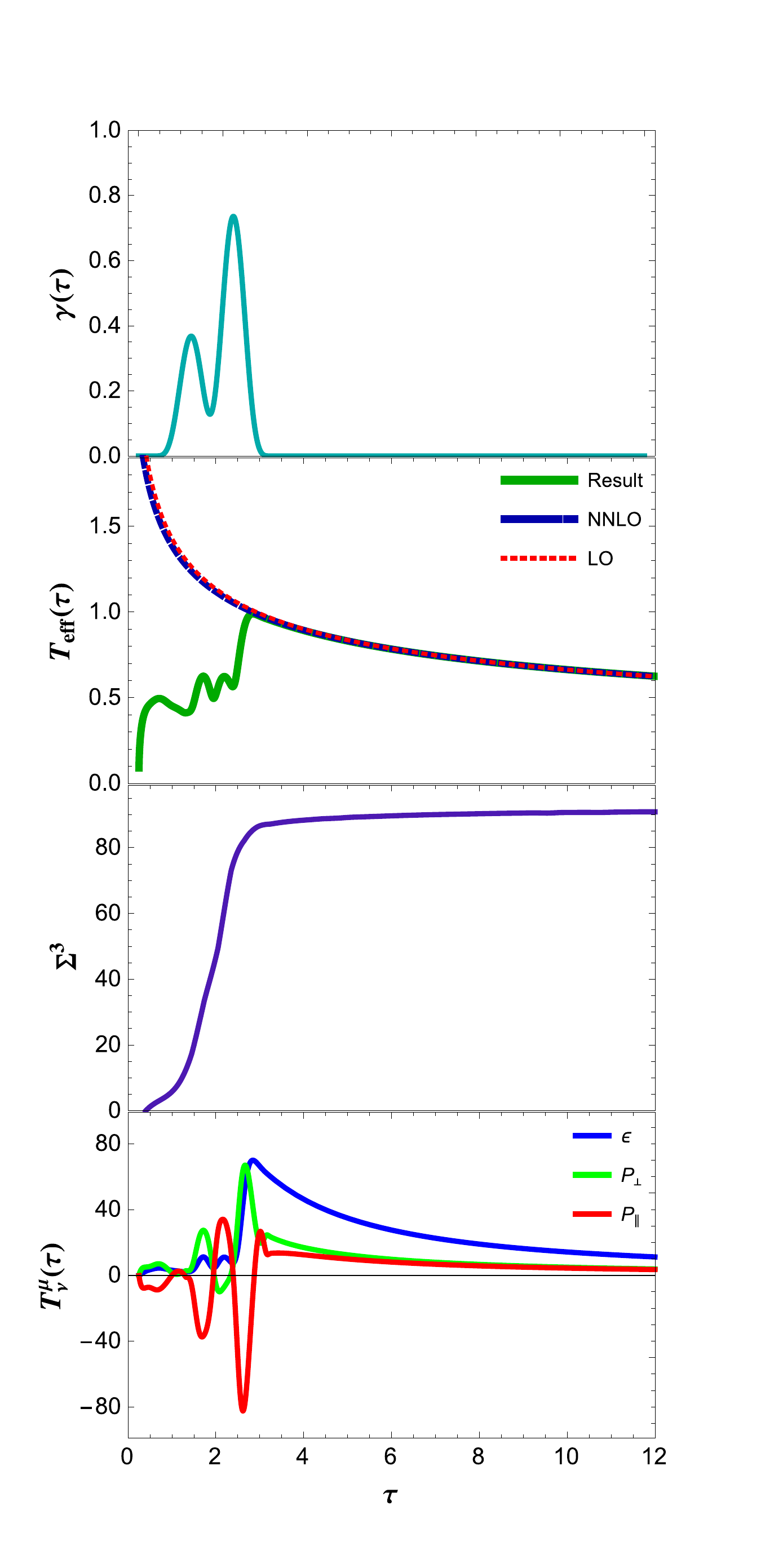} 
\end{tabular}
\vspace*{-0.5cm}
\caption{\small {\bf Model $\cal A$.} The  panels,  for  the profile $\gamma(\tau)$ with different time intervals between the pulses (left and right columns),  show (from top to bottom) the function $\gamma(\tau)$,    temperature $T_{eff}(\tau)$,   horizon area $\Sigma^3(r_h,\tau)$ per unit of rapidity,  and the three components  $\epsilon(\tau)$, $p_\perp(\tau)$ and $p_\parallel(\tau)$ of the stress-energy tensor $T^\mu_\nu$  (apart from the factor $\frac{N_c^2}{2 \pi^2}$\,). }\label{Column2p}
\eec
\end{figure}

In the case of two distant pulses, the horizon is formed as a two-step process which follows the sequence of the boundary distortion, as one can infer considering the computed function
$A(r,\tau)/r^2$ which is depicted in Fig.~\ref{density-due-picchi}. The outgoing radial null geodesics are clearly separated by a critical geodesic, the event horizon, which starts plunging back in the bulk  immediately after each pulse in the boundary. This can be even more clearly seen in the plot of the $\tau$ dependence of the effective temperature $T_{eff}(\tau)$ and in the plot of the area of the apparent horizon per unit of rapidity, $\Sigma(r_h,\tau)^3$, collected in Fig.~\ref{Column2p}. The temperature has two decreasing regimes, the first one after the first pulse, with the system starting relaxation   immediately after the end of the boundary distortion. Time intervals of about twice the duration of the pulse are very long with respect to the  relaxation time. The same phenomenon is observed  considering the function 
$\Sigma(r_h,\tau)^3$, which shows two time ranges in which it reaches a constant value, soon after the pulses. However, on such time scales, the components of the stress energy tensor have their own (computed) behavior, and no isotropization is observed after the first pulse.

On the other hand, for a distortion represented by two  close (nearly overlapping) pulses, no particular structure is observed before the end of  the boundary perturbation. The horizon area per unit of rapidity, $\Sigma(r_h,\tau)^3$, has a sharp increase during the distortion, and reaches the asymptotic large $\tau$ value $\Sigma(r_h,\tau)^3= \pi^3 \Lambda^2$ soon after the end of the distortion.  

\begin{figure}[thb!]
\bec
\begin{tabular}{ll}
\includegraphics[width = 0.4\textwidth]{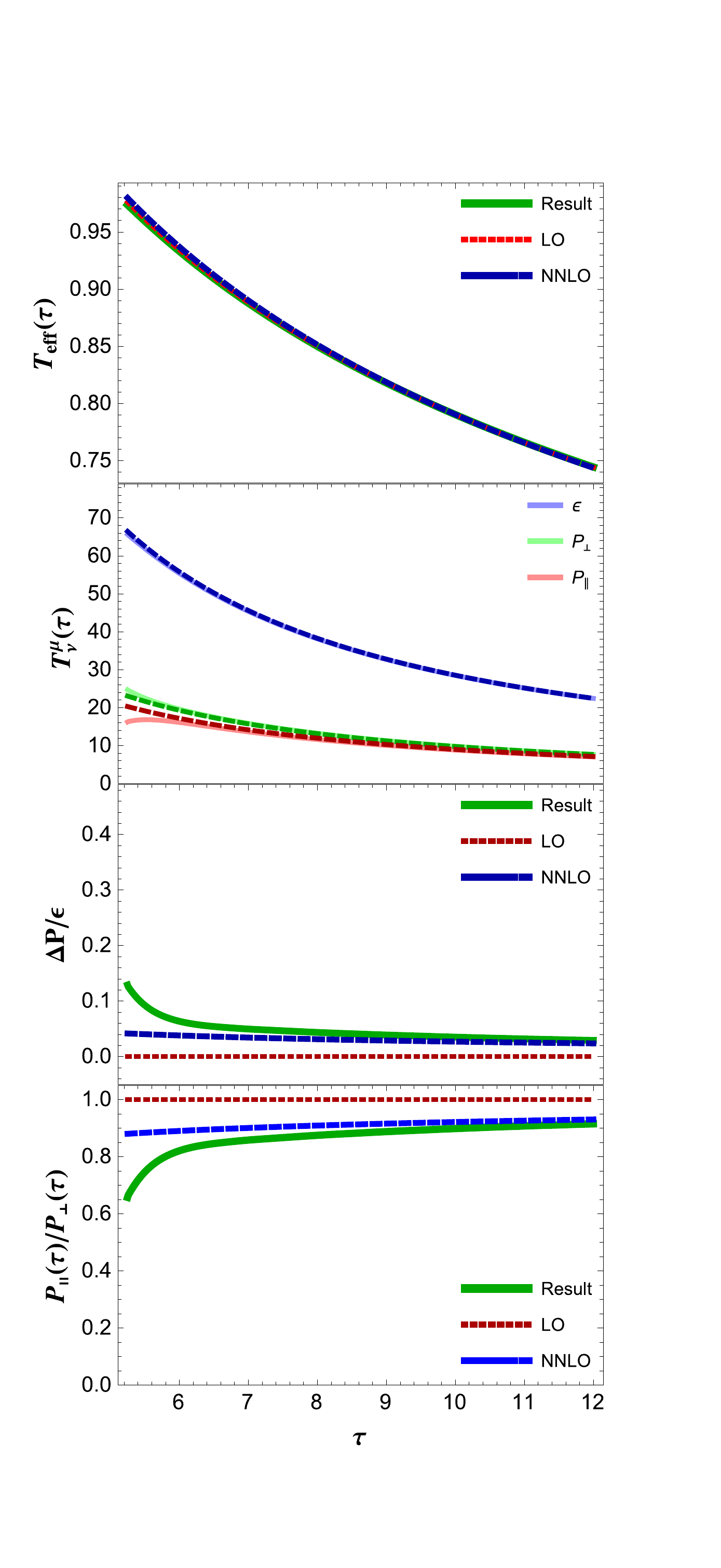}&\hspace*{-1.cm}
\includegraphics[width = 0.4\textwidth]{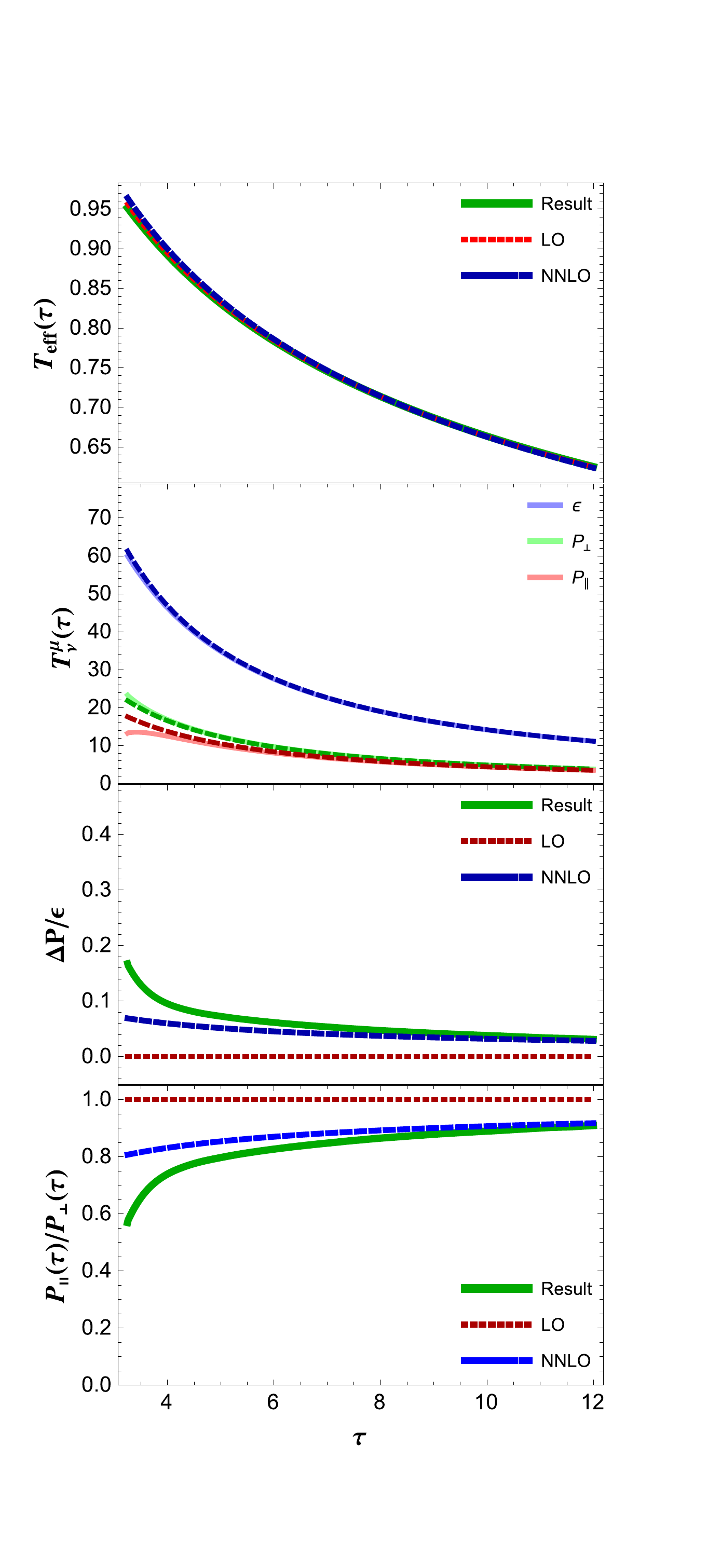}  
\end{tabular}
\eec
\vspace*{-1.5cm}
\caption{\small {\bf Model $\cal A$.} From top to bottom: temperature $T_{eff}(\tau)$,    components  $\epsilon(\tau)$, $p_\perp(\tau)$ and $p_\parallel(\tau)$ of the stress-energy tensor,
pressure anisotropy  $\Delta p/\epsilon=(p_\perp-p_\parallel)/\epsilon$ and ratio $p_\parallel/p_\perp$, computed for $\tau > \tau_f^{\cal A}$, for the boundary distortion with two distant (left) and two overlapping pulses (right panels).  The short and long dashed lines correspond to the hydrodynamic result and to the NNLO result in the $1/\tau$ expansion. }\label{Column2p-a}
\end{figure}

The hydrodynamic $\tau-$dependence of $T_{eff}(\tau)$ and of the energy density $\epsilon(\tau)$, Eqs.~(\ref{Teff1}) and (\ref{hydroeps}), is reached immediately after the end of the distortion of the boundary $\tau_f^A$, with values of the parameter $\Lambda$ which are different  for the two cases of different time intervals between the pulses. The values of $\Lambda$ obtained from  $T_{eff}(\tau)$  are collected in Table \ref{tab}. The temperature can be obtained from the condition  $\dot \Sigma(r_h,\tau)=0$, or from the energy density  Eq.(\ref{Teff}),  with numerical results differing at the level of $2 \%$. 
For a quantitative discussion, a time   $\tau^*$ can be fixed imposing that the energy density $\epsilon(\tau^*)$ differs from the hydrodynamic value $\epsilon_H(\tau^*)$  in  \eqref{hydroeps}
by less than  $5 \%$:
\be
\Big|\frac{\epsilon(\tau^*)-\epsilon_H(\tau^*)}{\epsilon(\tau^*)}\Big|=0.05 \,. \label{taustar}
\ee
In both cases of distant or overlapping pulses,  $\tau^*$ essentially coincides with the end of the boundary metric distortion. 
From the plot of $T^\mu_\nu$ in Fig. \ref{Column2p-a},  we observe that at $\tau^*$ the pressure anisotropy is still sizeable, and the values of $\Delta p/\epsilon$ and $p_{||}/p_\perp$ are still much larger than the ones expected by viscous hydrodynamics. 
%As  pointed out, e.g.,  in \cite{Heller:2012je,Ruggieri:2013ova},
%we also find that equilibration and isotropization are different processes, the former referring to temperature behavior and the latter to the pressures. 
Therefore,  we define the "thermalization time" $\tau_p$ from the condition  

\be
\Big|\frac{p_{||}(\tau_p)/p_\perp(\tau_p)-(p_{||}(\tau_p)/p_\perp(\tau_p))_H}{p_{||}(\tau_p)/p_\perp(\tau_p)}\Big| =0.05 \,\,\, , \label{taup}
\ee
 where $(p_{||}(\tau_p)/p_\perp(\tau_p))_H$ is given by \eqref{pressure-ratio}.
Quantitatively, from the condition that $p_\parallel/p_\perp$ differs by less than $5 \%$ from the asymptotic NNLO expression, we can set 
 $\tau_p=6.8$ for  two distant pulses. The  difference $\tau_p-\tau^*$ can be expressed in physical units if one scale in the system is set. Imposing 
$T_{eff}(\tau^*)=500$ MeV corresponds to  $\tau_p-\tau^*\simeq 0.60$~fm/c, which indicates  the elapsed time between  the end of the pulse and the restoration of  the hydrodynamical regime.
% for the plasma to reach the condition of pressure isotropy.
For two overlapping pulses, the time difference is $\tau_p-\tau^*\simeq 1.03$~fm/c. The numerical values are collected in Table \ref{tab}. Notice that
$\tau^*$ and $\tau_p$ differ in general,
although  Eqs.~\eqref{eq-ppar} and  \eqref{eq-pperp} hold. Indeed,
at a generic $\tau$  the relation between energy density and its asymptotic limit can be written as:
\be
\epsilon (\tau)= \xi(\tau) \epsilon_{H}(\tau) \label{rel-eps}
\,\,,
\ee
with $\xi(\tau^*)$ such that \eqref{taustar} is verified. For,  e.g.,  the longitudinal pressure,  using \eqref{eq-ppar} and (\ref{rel-eps})  one has 
$p_\parallel (\tau^*)=\xi(\tau^*) \, p_{\parallel , \, H}(\tau^*) -\tau^* \,\xi^\prime(\tau^*) \epsilon_{H}(\tau^*)$.
Therefore,  due to the  last term, the value $\tau_p$  verifying  \eqref{taup}  does not necessarily coincide with $\tau^*$.

\vspace*{0.5cm}

\begin{figure}[th!]
\begin{tabular}{ll}
\includegraphics[width = 0.45\textwidth]{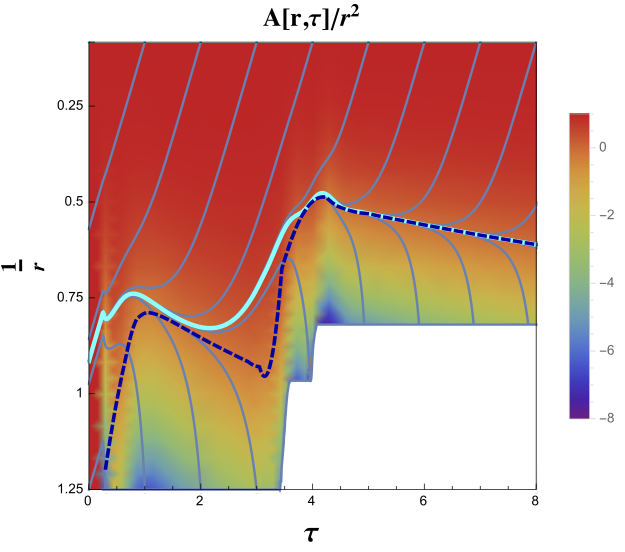}&\hspace*{-0.5cm}
\includegraphics[width = 0.5\textwidth]{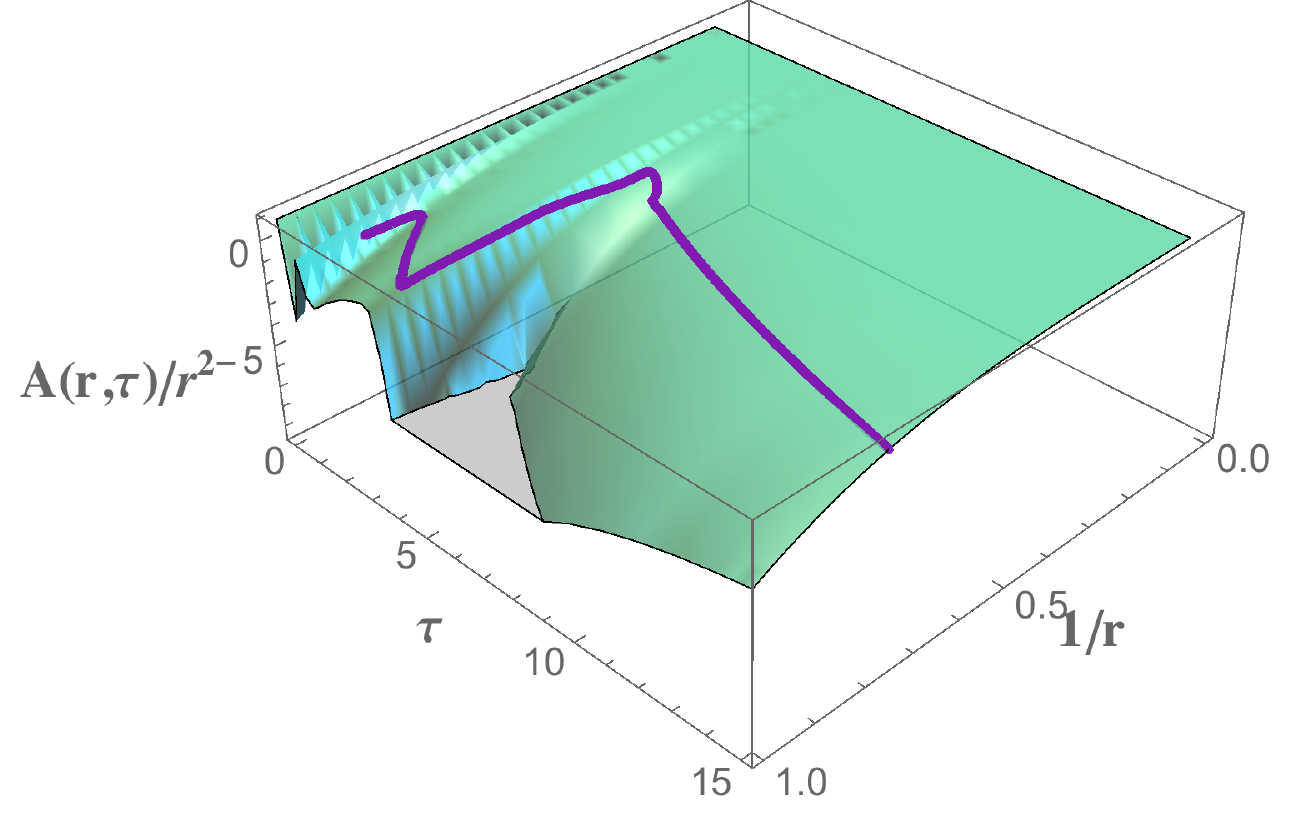}
\end{tabular}
\caption{\small {\bf Model $\cal B$.}  Function $A(r,\tau)/r^2$, vs $\tau$ and $1/r$, outgoing radial null geodesics,  event horizon and apparent horizon. Lines and colours have the same meaning as  in Fig.~\ref{density-due-picchi}. }\label{density-grad-picco}
\end{figure}

\vspace*{1cm}

\noindent {\bf Model $\cal B$.} 

\begin{figure}[t!]
\bec
\includegraphics[width = 0.4\textwidth]{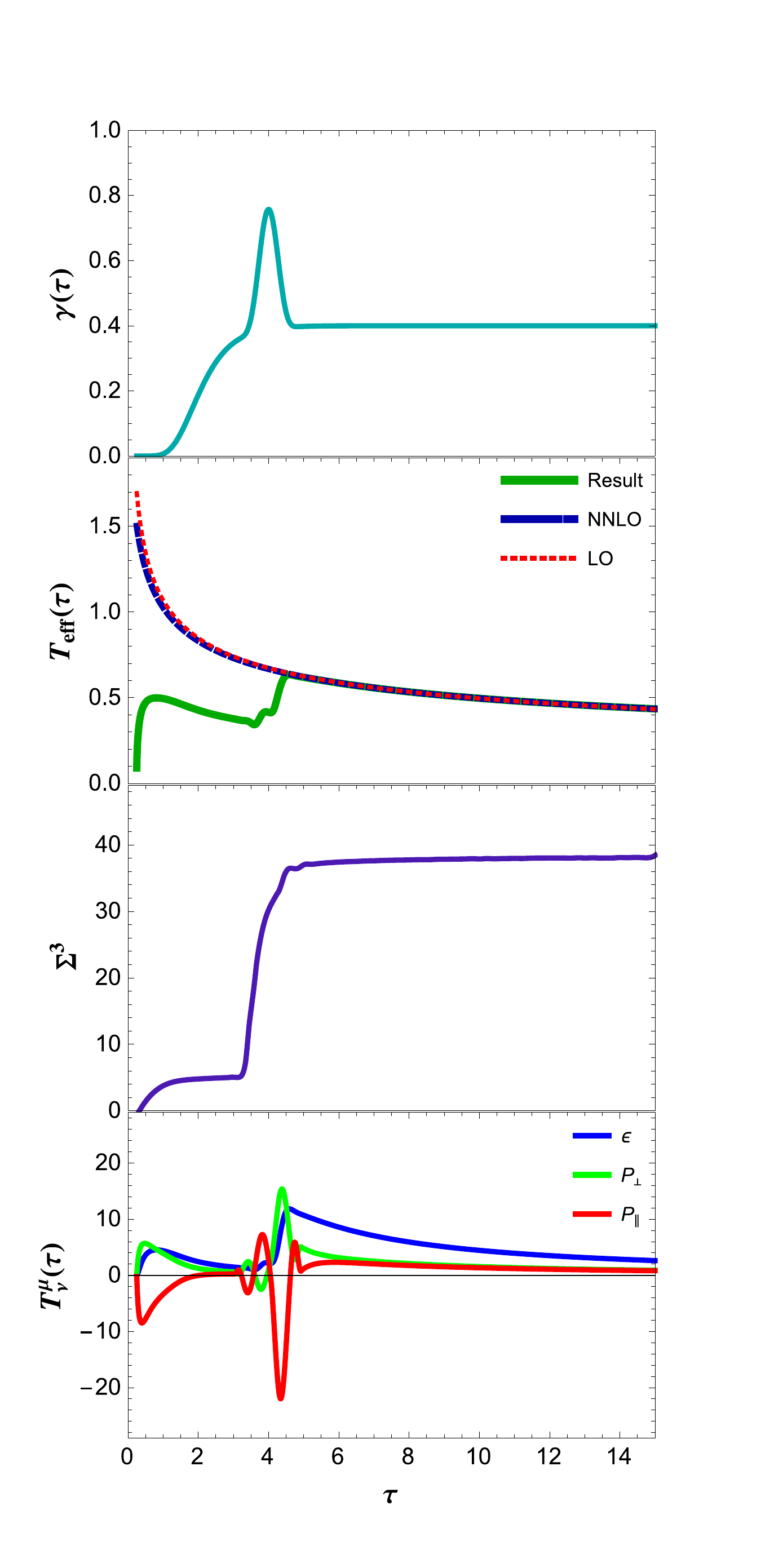}
\eec
\vspace*{-0.5cm}
\caption{\small {\bf Model $\cal B$.}  The  panels show (from top to bottom) the profile $\gamma(\tau)$, temperature $T_{eff}(\tau)$,   horizon area  per unit of rapidity $\Sigma^3(r_h,\tau)$,  and the three components  $\epsilon(\tau)$, $p_\perp(\tau)$ and $p_\parallel(\tau)$ of $T^\mu_\nu$. }\label{ColumnGp}
\end{figure}

\begin{figure}[t!]
\bec
\includegraphics[width = 0.4\textwidth]{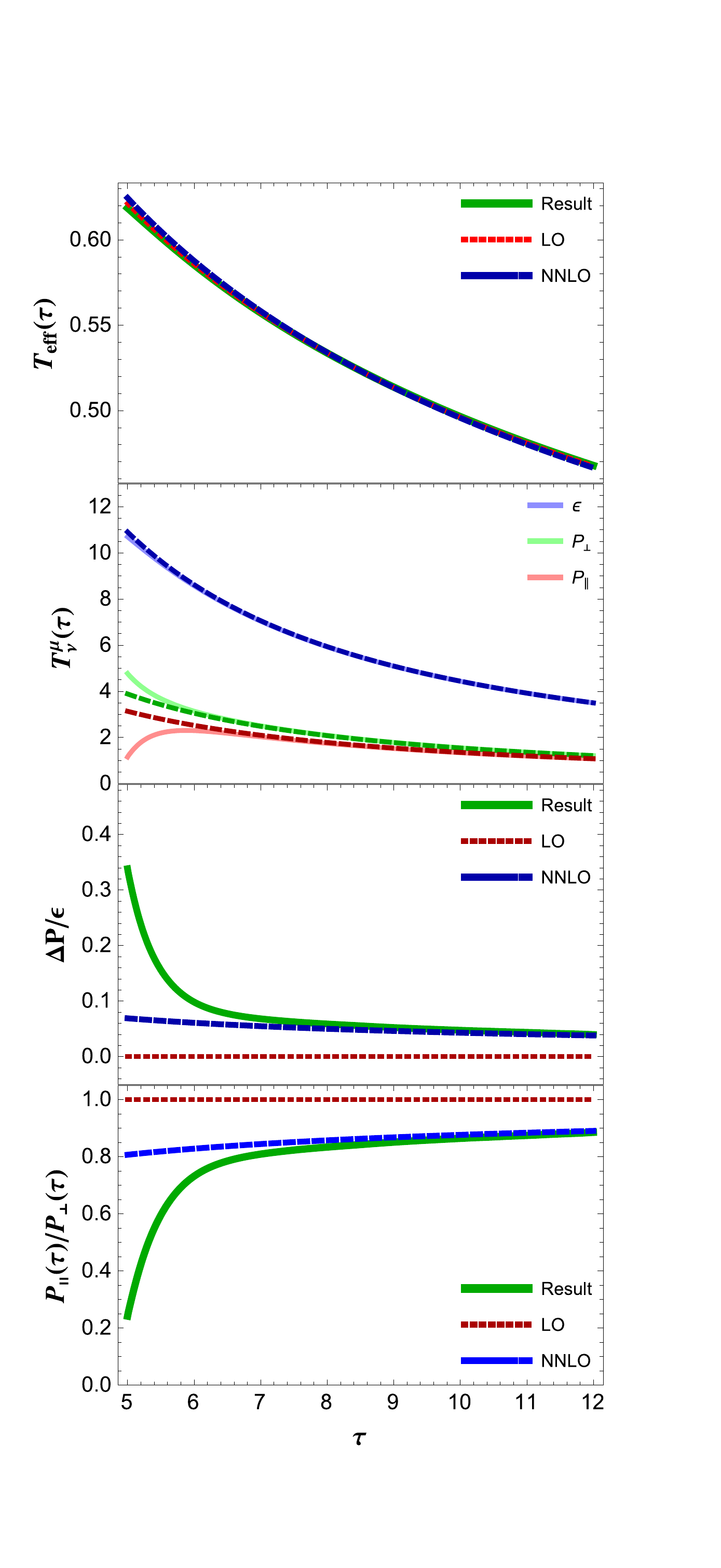} 
\eec
\vspace*{-1cm}
\caption{\small {\bf Model $\cal B$.}  From top to bottom: temperature $T_{eff}(\tau)$,    components  $\epsilon(\tau)$, $p_\perp(\tau)$ and $p_\parallel(\tau)$ of the stress-energy tensor,
pressure anisotropy  $\Delta p/\epsilon=(p_\perp-p_\parallel)/\epsilon$ and ratio $p_\parallel/p_\perp$, computed for $\tau > \tau_f^{\cal B}$.  The dashed lines correspond to the NNLO expressions.}\label{ColumnGp-a}
\end{figure}

The purpose of this model is to investigate a boundary distortion characterized by two largely different time scales. The horizon is formed immediately after the boundary deformation is switched on, as depicted in Fig.~\ref{density-grad-picco}, then starts plugging down in the bulk
until the short pulse  becomes active. This can be seen clearly in Fig.~\ref{ColumnGp}, where the effective temperature and $\Sigma^3$ display a decreasing and constant regime, respectively, in the proper time interval $\tau_i < \tau < \tau_{0,1}$, after which a violent perturbation takes the system out of equilibrium. The time  $\tau^*$ can be fixed using \eqref{hydroeps}. For $\tau > \tau^* \simeq \tau_f^{\cal B} $ the determination of  $\Lambda$ from $T_{eff}$ and the energy density $\epsilon$ gives $\Lambda=1.12$,  with a variation of  $3\%$ around this value. Thermalization  is reached at the time $\tau_p$ obtained by the condition \eqref{taup},  as one can infer 
considering  the observables  depicted in Figs.~\ref{ColumnGp} and \ref{ColumnGp-a}. Setting the scale so that the  temperature is  $T_{eff}(\tau^*)=500$ MeV, we find  $\tau_p-\tau^*\simeq 0.42$~fm/c.

\vspace*{0.5cm}
\noindent {\bf Model $\cal C$.}  

\begin{figure}[th!]
\begin{center}
\begin{tabular}{ll}
\includegraphics[width = 0.45\textwidth]{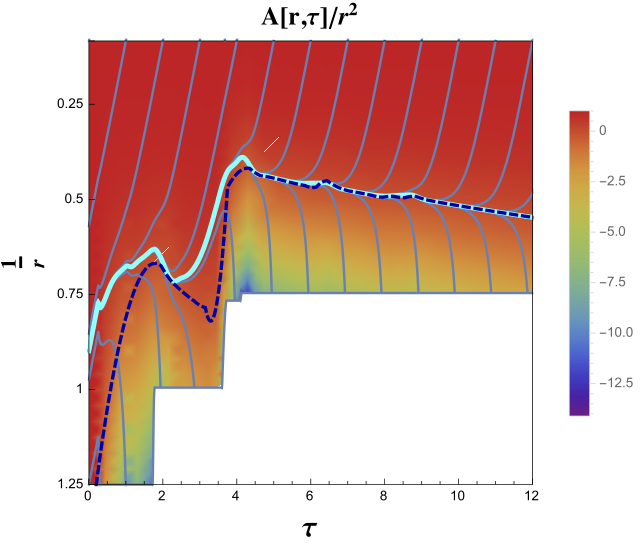}&\hspace*{-0.5cm}
\includegraphics[width = 0.5\textwidth]{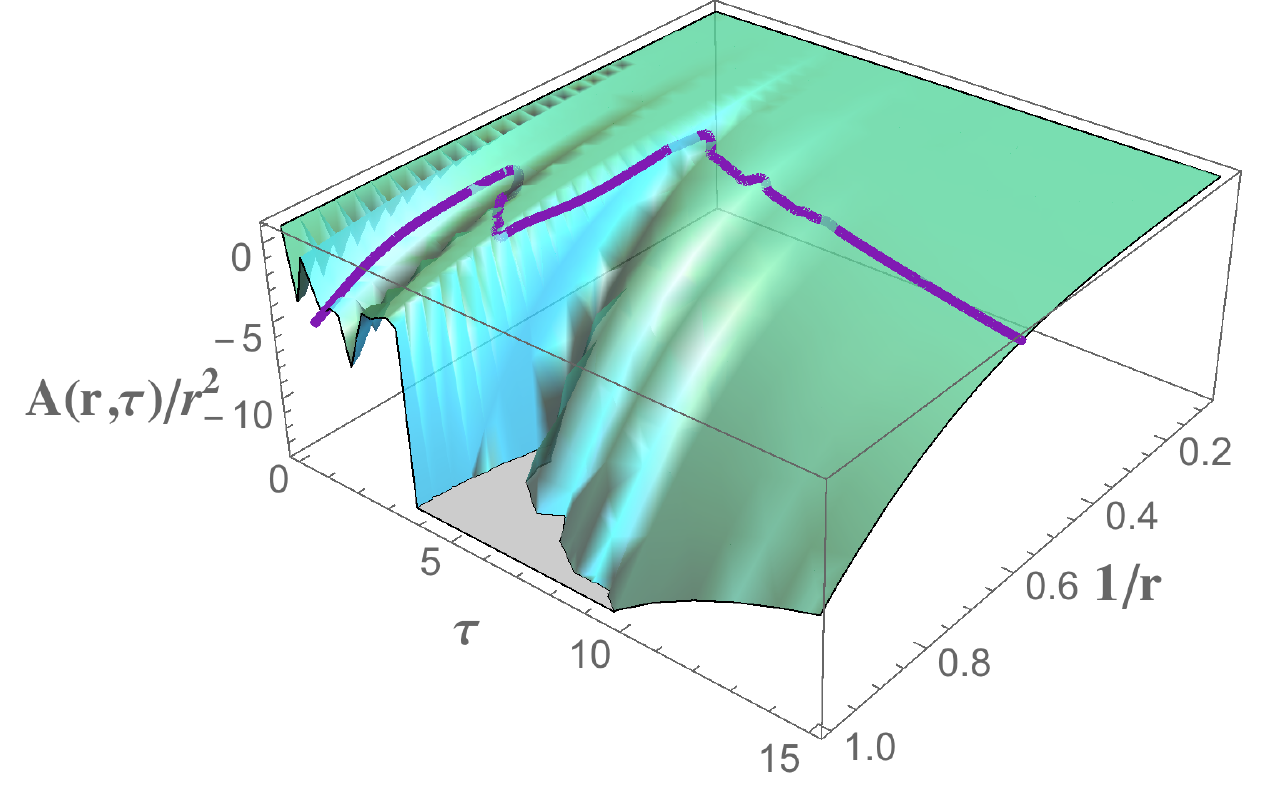}
\end{tabular}
\caption{\small {\bf Model $\cal C$.} Function $A(r,\tau)/r^2$, vs $\tau$ and $1/r$, outgoing radial null geodesics,  events horizon and apparent horizon. The lines and colors are indicated  as in Fig.~\ref{density-due-picchi}. }\label{density-grad-4picchi}
\end{center}
\end{figure}

\begin{figure}[t]
\bec
\includegraphics[width = 0.45\textwidth]{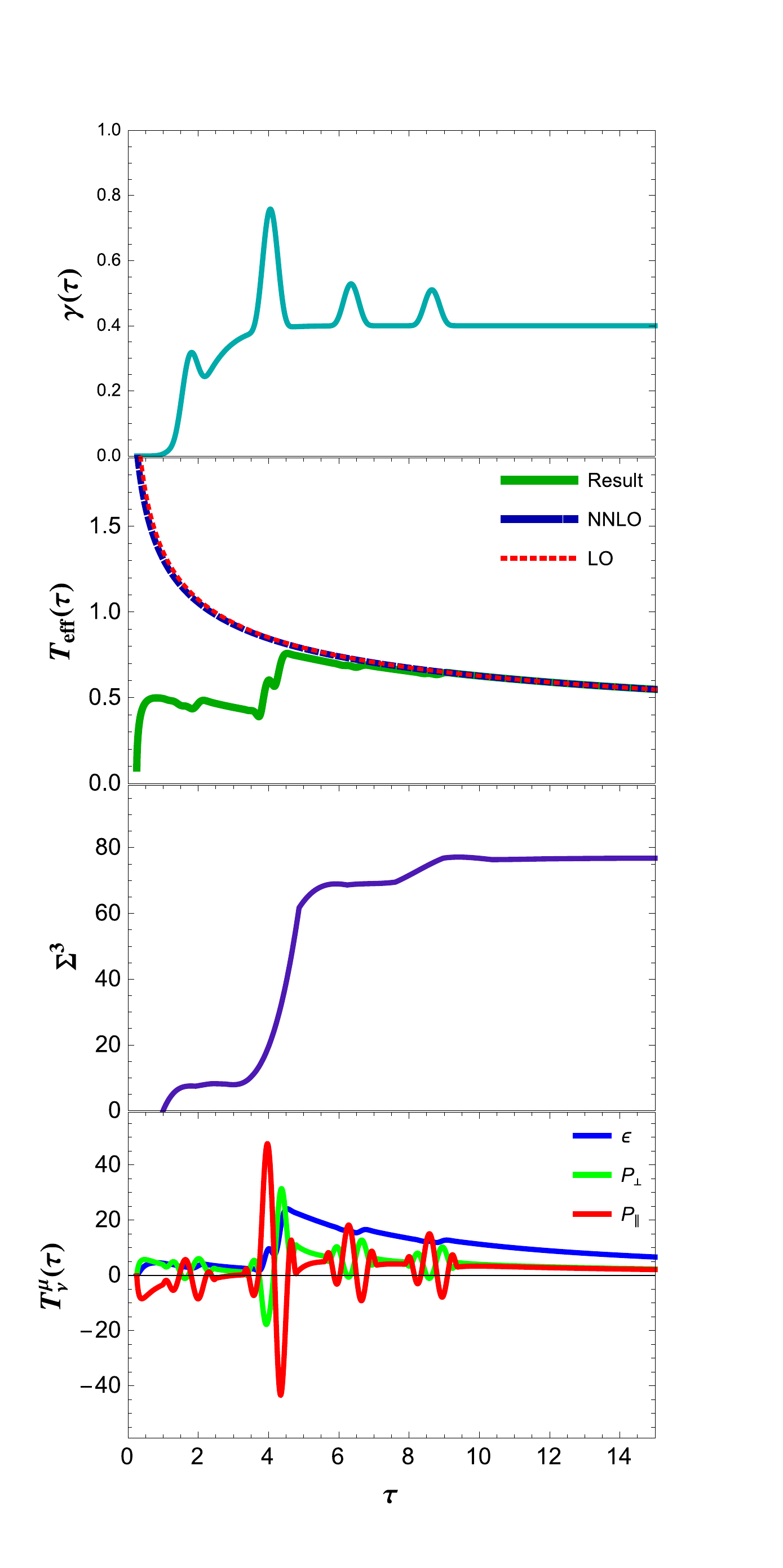}
\eec
\vspace*{-1cm}
\caption{\small {\bf Model $\cal C$.}  From top to bottom:  profile $\gamma(\tau)$,  temperature $T_{eff}(\tau)$,   horizon area  per unit of rapidity $\Sigma^3(r_h,\tau)$,  and the three components  $\epsilon(\tau)$, $p_\perp(\tau)$ and $p_\parallel(\tau)$ of $T^\mu_\nu$. }\label{Column4p}
\end{figure}

\begin{figure}[t!]
\bec
\includegraphics[width = 0.45\textwidth]{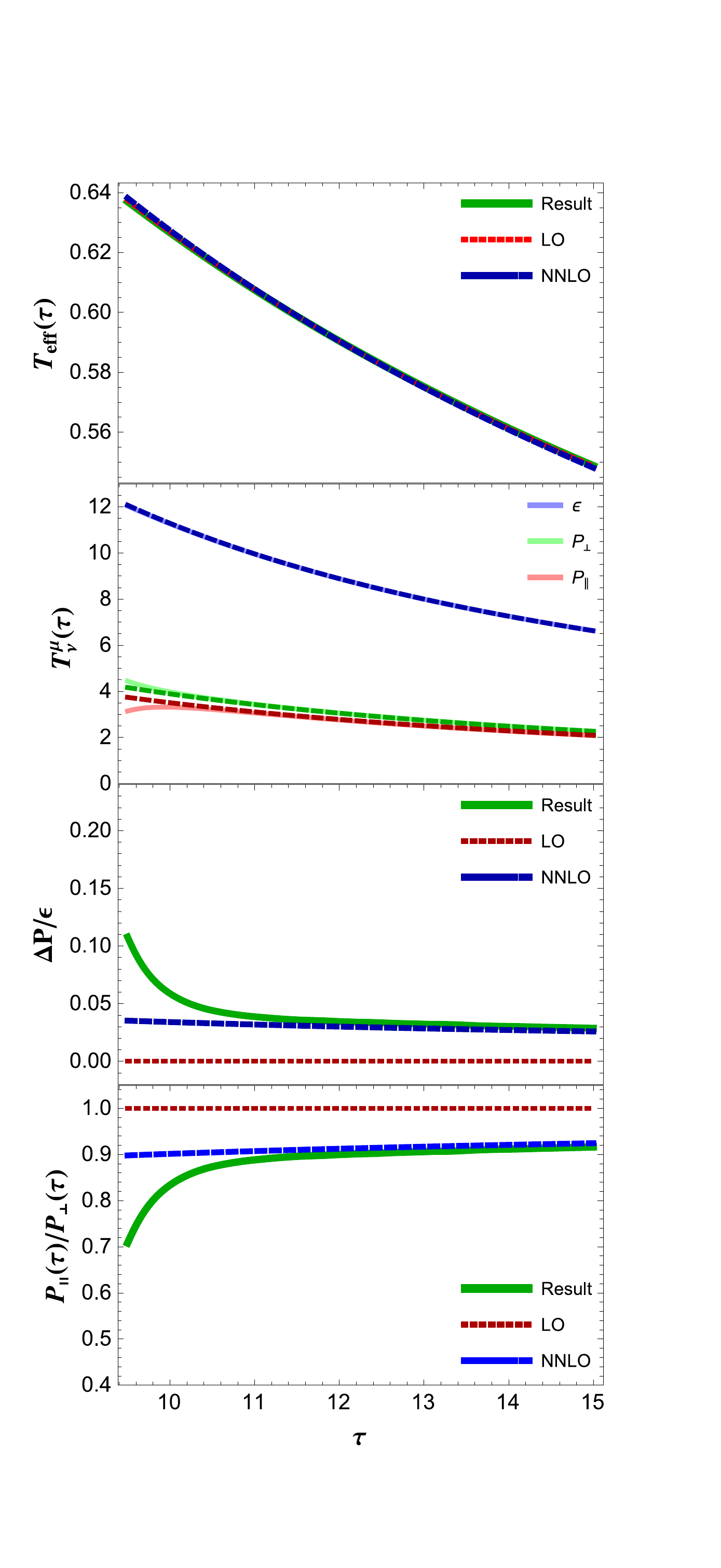}
\eec
\vspace*{-1cm}
\caption{\small {\bf Model $\cal C$.} From top to bottom: temperature $T_{eff}(\tau)$,    components  $\epsilon(\tau)$, $p_\perp(\tau)$ and $p_\parallel(\tau)$ of the stress-energy tensor,
pressure anisotropy  $\Delta p/\epsilon=(p_\perp-p_\parallel)/\epsilon$ and ratio $p_\parallel/p_\perp$, computed for $\tau > \tau_f^{\cal B}$.   The curves have the same meaning as in Fig.~\ref{ColumnGp-a}. }\label{Column4p-a}
\end{figure}

This model is constructed with the aim of representing a situation close to the physical system during the scattering process, with two different time scales and several pulses of various number and intensity. Immediately after switching on  the boundary
deformation, the horizon is formed in the bulk geometry and follows the profile of the distortion. Even in short time intervals between the pulses, the horizon starts to plung in the bulk, with a decreasing temperature and a saturation of $\Sigma^3$. This can be inferred considering the computed function $A/r^2$ depicted in Fig.~\ref{density-grad-4picchi} and, in details, studying the  $\tau$ dependence of the effective temperature and horizon area per unit rapidity  in Fig.~\ref{Column4p}. 
$\Sigma^3(r_h, \tau)$ displays a step-like behavior in proper time, closely following the distortion $\gamma(\tau)$, and reaches a constant value from $\tau=\tau_f^{\cal C}$ on. On the contrary, the various components of the stress-energy tensor  have  structures which become regular only for $\tau>\tau_f^{\cal C}$, after the last pulse.  As one can observe analyzing  the results in Fig.~\ref{Column4p-a}, the energy density reaches the  NNLO hydrodynamic behavior  at $\tau=\tau^*\simeq \tau_f^{\cal C}$, while pressure anisotropy persists for longer times. The parameter $\Lambda$, obtained from different observables, takes the value $\Lambda= 1.59$, with a variation similar to the one in models $\cal A$ and $\cal B$.  The pressure anisotropy $\Delta p/\epsilon$ and ratio $p_\parallel/p_\perp$ set the value $\tau_p$ of proper  time, as one can infer considering the results in Fig.~\ref{Column4p-a};  setting  $T_{eff}(\tau^*)=500$ MeV,  we obtain  $\tau_p-\tau^*\simeq 0.2$~fm/c.

\begin{table}[h]
\bec
\begin{tabular}{| c c c c c |}
\hline
model& $\tau^*$ & $\tau_p$ & $\Lambda $ & $\Delta \tau= \tau_p-\tau^*$ (fm/c)\\ \hline 
 ${\cal A}~(1)$ & 5.25 & 6.8 & 2.25 & 0.60 \\ 
 ${\cal A}~(2)$ & 3.25 & 6.0 & 1.73 & 1.03 \\ 
 ${\cal B}$ & 5 & 6.74 & 1.12& 0.42 \\ 
 ${\cal C}$ & 9.45 & 10.24 & 1.59 & 0.20 \\ \hline 
  \end{tabular}
\eec
\caption{Numerical values of the relevant quantities for  models $\cal A$, $\cal B$ and $\cal C$ of boundary distortion.}\label{tab}
\end{table}

\section{Conclusions}

We have investigated the effects of different types of distortions of the boundary metric, in a boost-invariant setup. The quenches are introduced as a way to take the boundary theory out-of-equilibrium,
and the distortion profiles are used to describe processes with different time scales  and intensities. As a common feature, we  observe a rapid formation of the horizon in the bulk metric, with the possibility of defining an effective
$\tau-$dependent temperature.  In coincidence with the end of each main distortion,  the  relaxation starts with the horizon plunging in the bulk. We have seen how sequences of distortions break the relaxation process.  
 We find that for all the considered distortion profiles, 
 $T_{eff}(\tau)$  starts to follow a viscous hydrodynamic expression  as soon as the quench is switched off. At this time the  pressures are different,  and evolve towards a common value which is reached at a later time.  Setting $T_{eff}(\tau^*)=500$ MeV,  the elapsed time before the restoration of the hydrodynamic regime  is  always a fraction of fm/c. Although the system described by the holographic approach  is in several respects different from the real QCD system, and  boundary sourcing   a quite abstract representation  of the heavy ion collision process, the obtained results  allow us to argue
what can be  expected in  realistic  situations.

\vspace*{0.5cm}
\noindent {\bf Acknowledgments.}\\
We thank Giuseppe Eugenio Bruno and Marco Ruggieri for  discussion. LB and PC thank  the Galileo Galilei Institute for Theoretical Physics for the hospitality and the INFN for partial support during the completion of this work.

\appendix
\section{Boundary stress-energy  tensor}\label{appA}
 For the sake of completeness, we briefly describe the main steps of  the calculation of the  boundary stress-energy tensor  $T^\mu_\nu$ leading to Eqs.(\ref{eps-a4})-(\ref{ppar-a4}). 
The $5D$ metric (\ref{AdS5}) can be written in  ADM form \cite{Arnowitt:1962hi},
 \begin{equation}\label{ADM}
ds^{2}=N^{2}dr^2+h_{\mu \nu} \left( dy^{\mu}+\mathcal{N}^{\mu}dr \right)
\left(dy^{\nu}+\mathcal{N}^{\nu}dr \right)
\end{equation} 
in terms of the induced metric  $h_{\mu \nu}$, of the lapse function
 $N$ and  of the shift vector field $\mathcal{N}^{\mu}$. 
 The stress tensor for the space foliation 
$\mathcal{M}_{r}$ obtained at constant $r$,    is given by
\begin{equation} \label{Ttildeab}
\left. \tilde{T}^{\mu \nu} \right|_{r}=\frac{2}{\sqrt{|h|}}
\left.\frac{\delta S}{\delta h_{\mu \nu}}\right|_{r},
\end{equation}
with $h\equiv det \left(h_{\mu \nu}\right)$ and $S$ the gravitational
action.  The boundary stress energy tensor  is obtained from the limit:
\begin{equation} \label{Tab implicit}
T^{\mu \nu}=\left.\lim_{r\to\infty} r^{2}\tilde{T}^{\mu \nu}\right|_{r}.
\end{equation}

The action $S$ in \eqref{Ttildeab}  includes as a counterterm  a local
functional $S_{ct}$ of  $h_{\mu \nu}$, whose contribution
to $\tilde{T}^{\mu \nu}$ regularizes the divergences at $r\to\infty$
\cite{Balasubramanian:1999re},\cite{Kinoshita:2008dq}.  As a consequence, 
the stress-energy tensor is  written as
\begin{equation} \label{Tabres}
T^{\mu \nu}=\lim_{r\to\infty}\frac{N_{c}^{2}}{4\pi^{2}} \, r^{2} \left(K^{\mu \nu}-
Kh^{\mu \nu}-3h^{\mu \nu}+\frac{1}{2}{\,}^{(4)}G^{\mu \nu}-\sigma^{\mu \nu}\frac{1}{r^2}\log
r \right), 
\end{equation}
with $K^{\mu \nu}$  the extrinsic curvature,
${\,}^{(4)}G^{\mu \nu}$ the boundary Einstein tensor
with  curvature tensors defined with respect to 
$h_{\mu \nu}$. The counterterms
$-3h^{\mu \nu}+\frac{1}{2}{\,}^{(4)}G^{\mu \nu}$
 cancel the powers from $r^{2}$ to $r^{-1}$  from the first two terms in (\ref{Tabres}),
 and introduce terms of order $r^{-2}$ which
contribute to the final result \eqref{Tabres} \cite{Balasubramanian:1999re}. The last term
proportional to  $\sigma^{\mu \nu}$, with in our case
\begin{eqnarray}
\sigma_{\mu \nu} &=& diag\left\{
-\frac{3}{2}\alpha_{4}(\tau),e^{\gamma(\tau)}\left[-\frac{1}{2}\alpha_{4}(\tau)+2\beta_{4}(\tau)
\right], \right. \nonumber \\
& & \left.
e^{\gamma(\tau)}\left[-\frac{1}{2}\alpha_{4}(\tau)+2\beta_{4}(\tau)
\right],e^{-2\gamma(\tau)}\tau^{2}\left[-\frac{1}{2}\alpha_{4}(\tau)-4\beta_{4}(\tau)\right]\right\} \,\,\, 
\end{eqnarray}
and  $\alpha_{4}(\tau)$ and $\beta_{4}(\tau)$  
in (\ref{ini2})-(\ref{ini1}), cancels the $\frac{1}{r^{2}}\log r$ contributions in (\ref{Tabres}).
The results for the components of the  stress-energy tensor  Eq.~(\ref{stress-en-tens})  are in Eqs.(\ref{eps-a4})-(\ref{ppar-a4}), with:
\bea
\tilde \epsilon_\gamma &=&
\frac{25}{144} \frac{\gamma'}{\tau^{3}}- \frac{\frac{65}{96}\gamma'^{2}+\frac{25}{144}\gamma''}{\tau^2}
+\frac{\frac{19}{32}\gamma'^{3}+\frac{1}{96}\gamma'(\tau)\gamma''-\frac{5}{48}\gamma^{(3)}}{\tau}
-\frac{57}{256}\gamma'^{4}-\frac{1}{8}\gamma''^{2}+\frac{5}{32}\gamma'\gamma^{(3)}\,, \nn \\
\label{Egamma}
\eea
\bea
\tilde p_{\perp\,\gamma} &=&-\frac{1}{6 \tau^4}+\frac{409}{576}\frac{\gamma'}{\tau^3}
-\frac{\frac{8}{9}\gamma'^{2}+\frac{587}{1728}\gamma''}{\tau^2}
+\frac{\frac{73}{192}\gamma'^{3}+\frac{145}{288}\gamma'(\tau)\gamma''-\frac{1}{288}\gamma^{(3)}}{\tau} \nn \\
&-&\frac{11}{256}\gamma'^{4}-\frac{21}{64}\gamma'^{2}\gamma''-\frac{1}{96}\gamma''^{2}+\frac{5}{96}\gamma'\gamma^{(3)}
+\frac{7}{192}\gamma^{(4)}\,, \label{PperpGamma}
\eea
\bea
\tilde p_{\|\,\gamma} &=&
\frac{1}{3\tau^{4}}-\frac{359}{288} \frac{\gamma'}{\tau^{3}}+\frac{\frac{49}{36}\gamma'^{2}+\frac{437}{864}\gamma''}{\tau^{2}}
-\frac{\frac{5}{12}\gamma'^{3}+\frac{233}{288}\gamma'\gamma''+\frac{7}{72}\gamma^{(3)}}{\tau} \nonumber \\
&-&\frac{11}{256}\gamma'^{4}+\frac{21}{32}\gamma'^{2}\gamma''-\frac{1}{96}\gamma''^{2}+\frac{5}{96}\gamma'\gamma^{(3)}
 -\frac{7}{96}\gamma^{(4)}\,. \label{PparGamma}
\eea
Energy density and pressures are hence affected by the variations of  $\gamma(\tau)$.
The above expressions coincide (after the correction of an overall  minus sign in $\tilde \epsilon_\gamma$)  with the ones in \cite{Chesler:2009cy} after the recognition of  terms due to the regularization scheme proportional to $\sigma_{\mu\nu}$,  related to
the coefficient  of the   $r^{-4}\log[r]$ term in  the large-$r$ expansion of the  metric.
The covariant conservation of the stress-energy tensor gives again  Eq.~(\ref{eq:a4b4}).

\section{Testing the numerical algorithm}\label{numerics}

\begin{figure}[t!]
\bec
\begin{tabular}{ll}
\includegraphics[width = 0.45\textwidth]{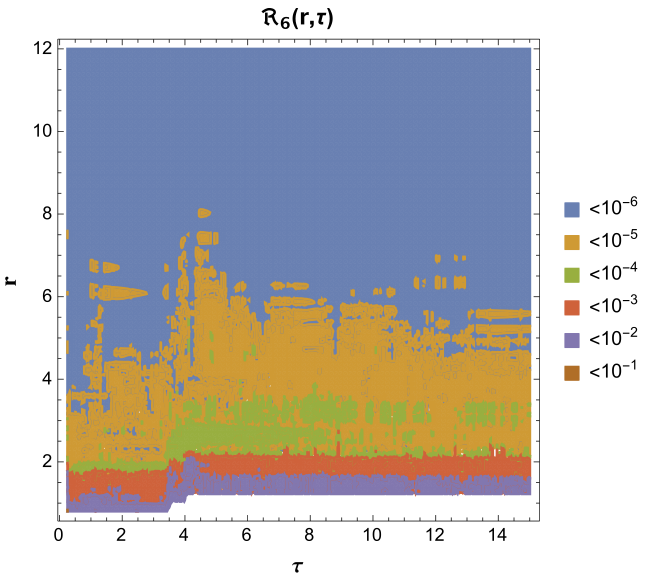}&\hspace*{-0.5cm}
\includegraphics[width = 0.45\textwidth]{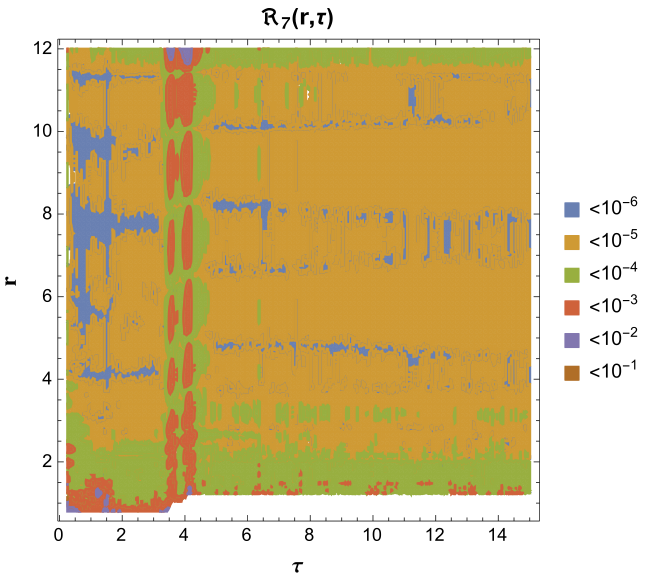}\\
\includegraphics[width = 0.45\textwidth]{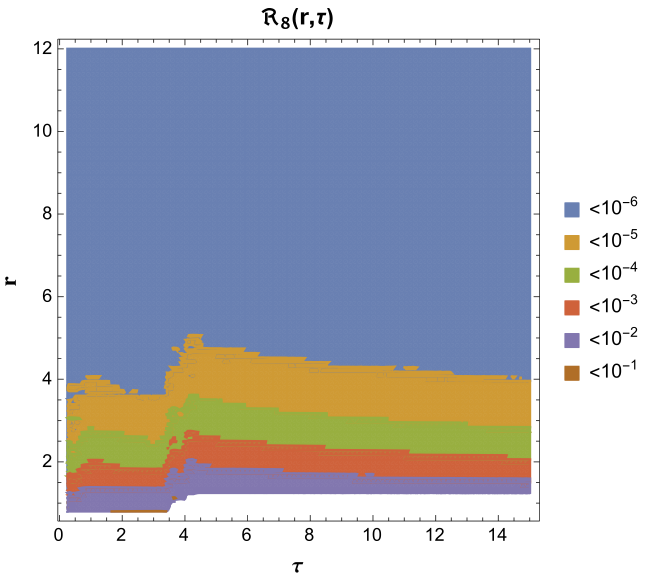}&\hspace*{-0.5cm}
\includegraphics[width = 0.45\textwidth]{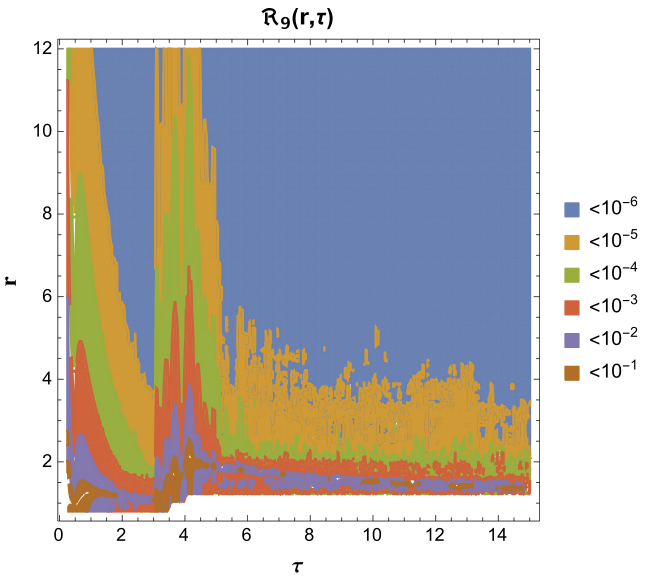}
\\
\includegraphics[width = 0.45\textwidth]{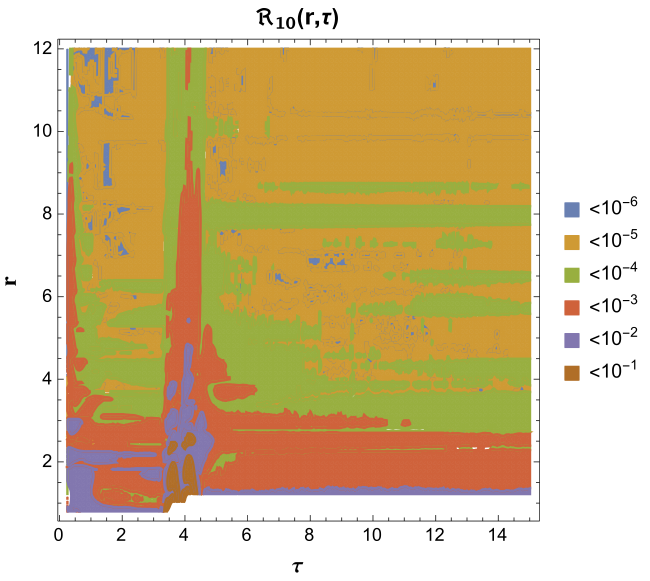}&\hspace*{-0.5cm}
\end{tabular}
\vspace*{-0.5cm}
\caption{\small  Ratios ${\cal R}_i(r,\tau)$ in the case of model $\cal B$. The white region corresponds to the excision in $r_{min}$ . }\label{Rplot}
\end{center}
\end{figure}

To quantify the accuracy of our numerical algorithm,  based on the Runge-Kutta method for the solution of the differential equations in the variable $r$, we monitor  the ratios 
 \bea
{\cal R}_6(r,\tau)&=&\frac{\Sigma ({\dot \Sigma})^\prime +2 \Sigma^\prime {\dot \Sigma}-2 \Sigma^2}{|\Sigma ({\dot \Sigma})^\prime| +2 |\Sigma^\prime {\dot \Sigma}|+2 | \Sigma^2|}  \label{normein1} \\
{\cal R}_7(r,\tau)&=& \frac{\Sigma ({\dot B})^\prime+\frac{3}{2} \left(\Sigma^\prime {\dot B}+B^\prime {\dot \Sigma}\right)}{|\Sigma ({\dot B})^\prime|+\frac{3}{2} \left(|\Sigma^\prime {\dot B}|+|B^\prime {\dot \Sigma}|\right)}  \label{normein2}\\
{\cal R}_8(r,\tau)&=& \frac{A^{\prime \prime} +3 B^\prime {\dot B} -12 \frac{\Sigma^\prime {\dot \Sigma} }{\Sigma^2}+4}{|A^{\prime \prime}| +3 |B^\prime {\dot B}| +|12 \frac{\Sigma^\prime {\dot \Sigma} }{\Sigma^2}|+4}\label{normein3}
\eea
\bea
{\cal R}_9(r,\tau)&= &\frac{{\ddot \Sigma}+\frac{1}{2} \left( {\dot B} ^2 \Sigma -A^\prime  {\dot \Sigma} \right) }{|{\ddot \Sigma}|+\frac{1}{2} \left(| {\dot B} ^2 \Sigma |+|A^\prime  {\dot \Sigma}| \right)}
\label{normein4} \\
{\cal R}_{10}(r,\tau)&=& \frac{\Sigma^{\prime \prime}+\frac{1}{2} B^{\prime 2} \Sigma}{|\Sigma^{\prime \prime}|+\frac{1}{2} B^{\prime 2} |\Sigma|} \label{normein5}
 \,\, \eea    
in the $(r,\,\tau)$ domain  behind the excision $r>r_{min}$.  The final results are shown in Fig.~\ref{Rplot} in the case of model $\cal B$.
\begin{figure}[t!]
\bec
\includegraphics[width = 0.45\textwidth]{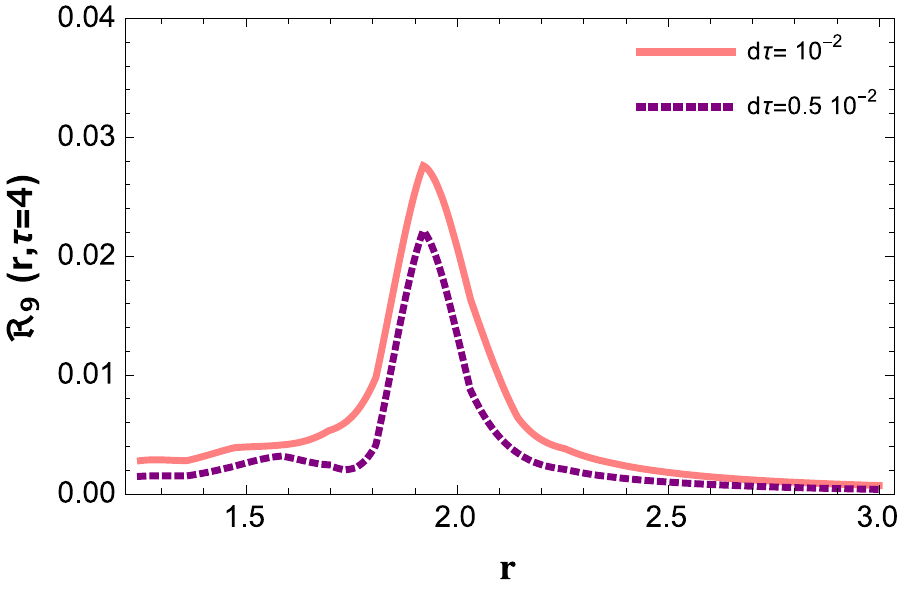}\\
\vspace*{-0.5cm}
\caption{\small Ratio ${\cal R}_9(r,\tau=4)$ computed in the range  $1.2 \leqslant r \leqslant 3.0$ for $d\tau=10^{-2}$ (pink, continuous curve) and  $d\tau=0.5 \cdot 10^{-2}$ (purple, dashed curve) in the case of model $\cal B$ . }\label{R9dtau}
\end{center}
\end{figure}
The resulting ratios ${\cal R}_6,\,{\cal R}_7,\,{\cal R}_8$ deviate from zero at a level smaller than ${\cal O}(10^{-4})$, but for a tiny  region close to the excision in the case of ${\cal R}_6$ and ${\cal R}_8$, and for a few spots  in the range   $3 \leqslant r \leqslant 5$ (where the source function $\gamma(\tau)$ is peaked) in the case of ${\cal R}_7$.
A similar result is obtained for ${\cal R}_{9}$ and ${\cal R}_{10}$, with larger deviations from zero:   this is a consequence, for ${\cal R}_{10}$, of the smallness of the two addendi in  (\ref{ein5}), and for ${\cal R}_{9}$ of the computation of ${\ddot \Sigma}=\partial_\tau {\dot \Sigma}+\frac{1}{2} A\,(\dot \Sigma)^\prime$ through  a discretized time derivative.
To monitor the time slicing, we compute ${\cal R}_9$ for two different time steps in the region $1.2 \leqslant r \leqslant 3.0$ at $\tau=4$, where there is the largest  deviation from zero.   Comparing the results  for  $d\tau= 10^{-2}$ and  $d \tau=0.5 \cdot 10^{-2}$,
 as shown in Fig. \ref{R9dtau}, one observes the decreasing of ${\cal R}_9$ by doubling  the grid points.

\bibliographystyle{JHEP}
\bibliography{our-ref}
\end{document}